# Energy and momentum of the surface plasmon-polariton supported by a thin metal film

A. Y. Bekshaev*[a], O. V. Angelsky[b]
[a]Physics Research Institute, I.I. Mechnikov National University, Odessa, Ukraine;
[b]Yuriy Fedkovych Chernivtsi National University, Chernivtsi, Ukraine
**bekshaev@onu.edu.ua*

**Abstract**

We study the energy and momentum of the surface plasmon-polariton (SPP) excited in a symmetric 3-layer "insulator-metal-insulator" structure, which is known to support the symmetric (S) mode with the negative group velocity as well as the antisymmetric (AS) mode with only positive energy flow. The electric and magnetic field vectors are calculated via both the phenomenological and the microscopic approach; the latter involves the hydrodynamic model accounting for the quantum statistical effects for the electron gas in metal. Explicit representation for the energy and momentum constituents in the dielectric and in the metal film are obtained, and the wavenumber dependences of the energy and momentum contributions for the whole SPP are analyzed numerically. The various energy and momentum constituents are classified with respect to their origin: "field" or "material", and the physical nature: orbital (canonical) and spin (Belinfante) momentum contributions. The pictures characteristic for the S and AS modes are systematically compared. The results can be useful for the studies and applications of the SPP-induced thin-film effects, in particular, for the charge and spin dynamics in thin-film plasmonic systems.

## 1. Introduction

Light fields with complex and highly developed spatial structure –"structured light" – attract significant attention due to their impressive abilities within optical nano-probing, precise optical manipulation and optical data processing [1–3]. The especial interest is directed to the surface plasmon-polariton (SPP) waves emerging near the interface between dielectric and conductive media [1–7]. Their unique properties, such as the "extraordinary" transverse spin and momentum [7–9], possibility of the special spin-momentum locking [9,10], nonreciprocity and unidirectional propagation [9–12] stimulate promising applications for the optical information techniques and emerging nanotechnologies.

The interface and the corresponding strong spatial inhomogeneity, as well as the dispersion (frequency dependence of the main electric and magnetic parameters), are the prerequisites for the SPP excitation and propagation. These conditions pose special problems in the theoretical description of the SPP fields, especially of their dynamical characteristics: energy, energy flow, momentum, angular momentum and their derivatives. Consistent solutions of these problems were recently proposed in a series of works [13–17], according to which the momentum and angular momentum of light in a dispersive medium can be described similarly to the known Brillouin's formula for the energy, by means of the special dispersive corrections of the medium permittivity and permeability [18,19]. The efficiency of this approach in application to the SPP was convincingly demonstrated [20,21]. Importantly, its results were confirmed by the microscopic calculations involving the direct consideration of the motion of electrons in the metal layer, based on the simplified Drude model [2,6,15–17] as well as on the hydrodynamic model [21–25] which takes the quantum pressure effects into account. Despite that the microscopic results are numerically close to those obtained by the phenomenological approach, the use of microscopic models is necessary for the meaningful classification of the momentum contributions with separation of the "field" and "material" manifestations, their specific nature and origination. Such a classification includes the momentum constituents "belonging" to the electromagnetic field 'per se' and a series of the momentum "blocks" emerging due to various rotational and translational components of the field-induced motion of charge carriers [21]. In this classification process, the physical background for the phenomenological terms, associated with the metal permittivity and the dispersion corrections, has been disclosed and scrutinized [15,17,21]. In turn, both the field and material momentum contributions are subjects of the canonical decomposition into the orbital (canonical) and spin (Belinfante) parts [15,16,21,26,27].

The single metal-insulator interface is the simplest but not the unique SPP-supporting structure. In many cases, the metal layer is thin and limited by the second boundary thus forming the 3-layer "insulator-metal-insulator" structures [6,28,29]. Then, the waves supported by each interface interact and form "hybrid" modes with different and sometimes extraordinary properties. For example, in the symmetric 3-layer structure (see Fig. 1 below), the two transverse-magnetic (TM) SPP modes can exist, and the mode with higher frequency (in which the instantaneous magnetic field is distributed symmetrically with respect to the middle plane) may show the negative group velocity and, accordingly, the energy flow directed oppositely to the wave propagation. According to [28], in this paper this mode is called "symmetric", or S mode. Another mode – antisymmetric (AS) with lower frequency – shows the more traditional behavior with decreasing but always positive group velocity.

Apparently, the specific picture of the negative energy flow is connected with certain peculiarities of the energy and momentum behavior in the S mode, and tracking of such connections is the aim of the present paper. In what follows, we briefly reproduce the main points of the phenomenological and microscopic description of the SPP electromagnetic field in the symmetric 3-layer system (Section 2) and apply them to the numerical calculations of the energy (Section 3) and momentum (Section 4) constituents depending on the SPP wavenumber. The results are presented as a collection of equations and graphs illustrating the energy and momentum "blocks" of different origins and physical nature. Their special features for the S and AS modes and underlying physical grounds are discussed in Section 5. The attention is focused on the "integral" energy and momentum of the SPP, characterizing integrated the "whole wave; the details of their spatial distributions over the film depth are discussed in the Supplementary Material (SM).

## 2. General description of the SPP in symmetric thin-film structure

We study the symmetric structure with a conductive layer 2 between two identical dielectric media 1 and 3 depicted in Fig. 1a. The structure of Fig. 1 supports monochromatic SPP waves with frequency $\omega$ corresponding to the vacuum wavenumber $k = \omega / c$ ($c$ is the light velocity). The SPP propagates along the $z$ direction, and all the field components are proportional to the phase factor $\exp(ik_s z)$ where $k_s$ is the SPP wavenumber. Let all media be non-magnetic with permeabilities $\mu_{1,2}$ = 1, and the dielectric layers $|x| > a$ are characterized by the constant real permittivity $\varepsilon_1$. The permittivity of the conductive film $|x| < a$ is described by a certain function $\varepsilon_2(\omega)$ whose form is postulated (in the phenomenological approach) or derived microscopically. For simplicity, the energy dissipation can be neglected in the first approximation [2,6,15–17,20,21], i.e. $\mathrm{Im}\,\varepsilon_2(\omega) = 0$.

### 2.1. Phenomenological approach

The electromagnetic field obeys the Maxwell equations [2,18,19]

$$\nabla \mathbf{H} = 0, \quad \mathbf{H} = \frac{1}{ik} \nabla \times \mathbf{E}, \tag{1}$$

$$\nabla \mathbf{E} = 0\,, \quad \mathbf{E} = -\frac{1}{ik\varepsilon}\nabla \times \mathbf{H} \tag{2}$$

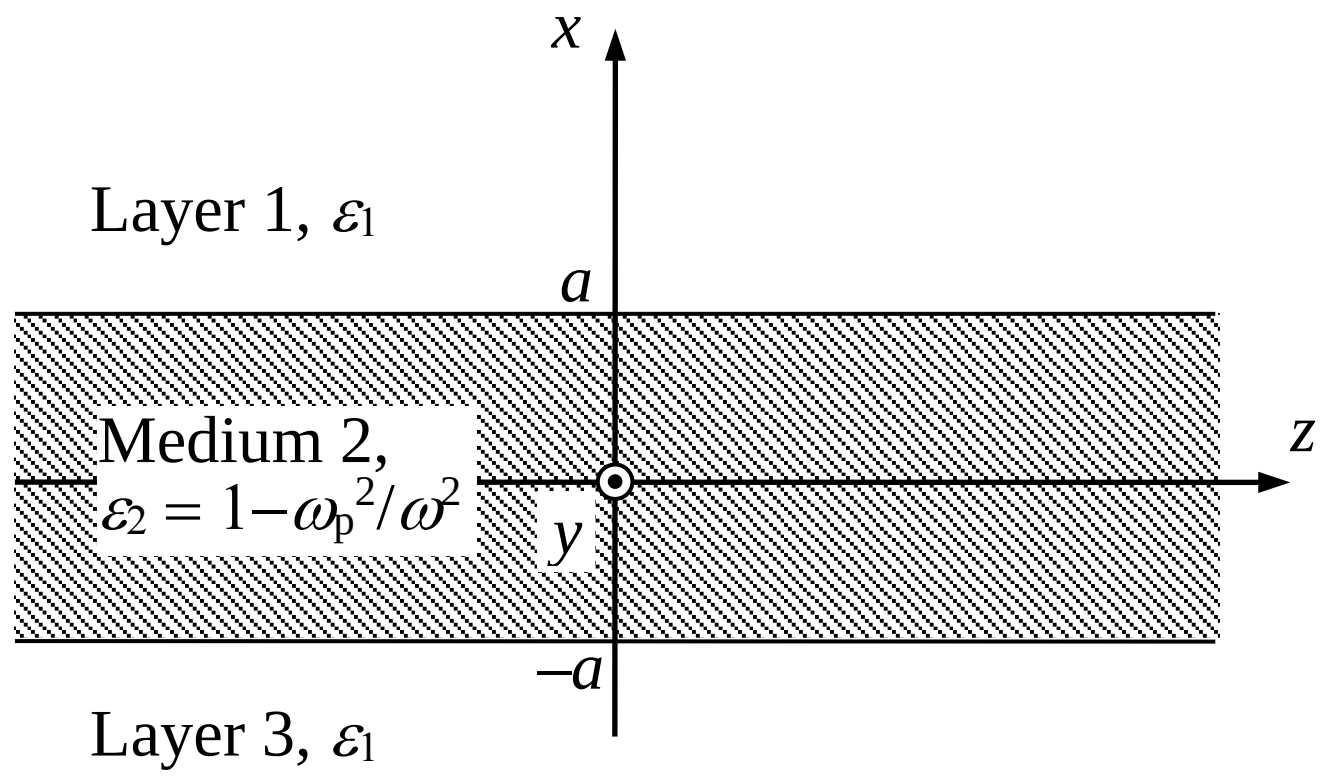

Fig. 1. Symmetric SPP-supporting structure: a metallic film 2 (width $2a$, frequency-dependent permittivity $\varepsilon_2(\omega)$) is enclosed between the dielectric layers 1 and 3 with the real frequency-independent permittivity $\varepsilon_1$. Axis $y$ is normal to the figure plane.

with **E** and **H** being the complex electric and magnetic fields. Generally, the SPP field is described by the TM solutions of Eqs. (1), (2) with the only non-zero magnetic-vector component $H_y$; the boundary conditions require continuity of $H_y$, $E_z$ and $\varepsilon E_x$. There are two sorts of such solutions: symmetric (S) and antisymmetric (AS) modes, regarding the symmetry or antisymmetry of the $H_y$ distribution with respect to the plane $x = 0$ [28,29]. For the S mode, the field expressions are obtained in the forms:

in the layers 1 and 3 ($|x| > a$)

$$H_y = A\frac{k}{k_s}\exp\left(-\kappa_1|x|\right)e^{ik_s z}\,,\;\; E_x = \frac{A}{\varepsilon_1}\exp\left(-\kappa_1|x|\right)e^{ik_s z}\,,\;\; E_z = -i\,\mathrm{sgn}\left(x\right)A\frac{\kappa_1}{k_s\varepsilon_1}\exp\left(-\kappa_1|x|\right)e^{ik_s z}\,; \tag{3}$$

in the film 2 ($-a < x < a$)

$$H_y = Ae^{-\kappa_1 a}\frac{k}{k_s}\frac{\cosh\kappa_2 x}{\cosh\kappa_2 a}e^{ik_s z}\,,\;\; E_x = \frac{A}{\varepsilon_2}e^{-\kappa_1 a}\frac{\cosh\kappa_2 x}{\cosh\kappa_2 a}e^{ik_s z}\,,\;\; E_z = iAe^{-\kappa_1 a}\frac{\kappa_2}{k_s\varepsilon_2}\frac{\sinh\kappa_2 x}{\cosh\kappa_2 a}e^{ik_s z}\;. \tag{4}$$

Here the conditions hold

$$\kappa_{1,2} = k_s^2 - k^2\varepsilon_{1,2}\,, \tag{5}$$

and the dispersion relation takes the form

$$\tanh\kappa_2 a = -\frac{\kappa_1\varepsilon_2}{\kappa_2\varepsilon_1}\,. \tag{6}$$

The results (3) – (6) are well known [28,29] and are reproduced here for the convenience of further references. Note that equality (6) is only possible when the right-hand side is positive, i.e. under the usual conditions of positive $\varepsilon_1$, negative sign of the film permittivity is required, $\varepsilon_2 < 0$. For the AS mode, in Eqs. (3) one should replace

$$\left(H_y,\, E_x,\, E_z\right) \to \mathrm{sgn}\left(x\right)\left(H_y,\, E_x,\, E_z\right) \tag{7}$$

and Eqs. (6) are valid with the substitutions

$$\cosh\kappa_2 x \rightleftarrows \sinh\kappa_2 x\,, \tag{8}$$

which, in particular, means that in the AS mode $H_y(x)$ and $E_x(x)$ are odd functions of $x$ while $E_z(x)$ is an even function (just opposite to the S-mode expressions (3), (4)). The dispersion relation for the AS mode differs from (6) by the fraction inversion in the right-hand side.

The electromagnetic field behavior, and the very existence of the propagating SPP modes, essentially depend on the function $\varepsilon_2(\omega)$; for simplicity and in view of further microscopic analysis, we accept the Drude model [2,6,15,21] for the conductive medium, in which

$$\varepsilon_2 = 1 - \frac{\omega_p^2}{\omega^2}, \quad \omega_p^2 = \frac{4\pi n_0 e^2}{m} \tag{9}$$

($\omega_p$ is the volume plasmon frequency). In this case, the dispersion law $\omega(k_s)$ derived from Eqs. (5), (6) is illustrated by the blue curve in Fig. 2 for the S mode and by the green curve for the AS one; with growing SPP wavenumber $k_s$ both curves asymptotically tend to the value $\omega_c = \omega_p \big/ \sqrt{1+\varepsilon_1}$, which corresponds to the cutoff frequency for a single-interface system [15,17,21]. Note that the S-mode SPP propagation with $\omega > \omega_c$ is possible (for $k_s > k_{s1}$); remarkably, at a certain point $k_{s2}$ the blue curve reaches a maximum, and the region of the negative group velocity ($d\omega/dk_s < 0$) exists at $k_s > k_{s2}$ (see also the brown curve in Fig. 3) whereas the AS mode shows a traditional behavior typical for the single-interface SPP (realized, e.g., when $a \to \infty$ in the model of Fig. 1).

### 2.2. Microscopic approach

In the microscopic approach [17,21], the dielectric layers 1 and 3 ($|x| > a$) are still characterized by the phenomenological permittivity $\varepsilon_1$, and Eqs. (1) – (3) are true. But the fields in the film 2 are determined with the explicit account for the motion of electrons. Therefore, in the region $|x| < a$, the Maxwell equations (1) still hold whereas Eqs. (2) are modified [23,21] to:

$$\nabla\mathbf{E} = 4\pi ne, \quad \mathbf{E} = \frac{i}{k}\nabla\times\mathbf{H} - i\frac{4\pi n_0 e}{\omega}\mathbf{v}. \tag{10}$$

Here, the velocity of electrons $\mathbf{v}$ and the charge density $ne$ obey the hydrodynamic equation for the electron gas [22–25]

$$-i\omega n_0 m\mathbf{v} = n_0 e\mathbf{E} - m\beta^2\nabla n. \tag{11}$$

where $n_0$ is the equilibrium electron density whose charge is compensated by the “background” charge of positive ions, $n$ is the non-equilibrium “excess”, and the quantum statistical effects are included via the coefficient $\beta^2 = (3/5)v_F^2$ involving the Fermi velocity of electrons $v_F$. In this model, the “background” permittivity of the medium 2 equals to 1, and in the boundary conditions, continuity of $H_y$, $E_z$ and $\varepsilon E_x$ is supplemented by requirement $v_x = 0$ at $x = \pm a$; negligence of the energy dissipation is expressed by the absence of the collision terms in Eq. (11) [22,24]. Results are as follows.

Instead of (6), for the dispersion relation we get

$$\tanh\kappa_2 a = -(1-\eta)\frac{\kappa_1}{\kappa_2\varepsilon_1} + \frac{k_s^2}{\kappa_2}\frac{\eta}{\gamma}\tanh\gamma a, \tag{12}$$

where the new parameter $\gamma$ appears due to the additional quantum pressure described by the last term of Eq. (11):

$$\gamma^2 - k_s^2 = -\frac{\omega^2}{\beta^2}(1-\eta), \quad \eta = \frac{\omega_c^2}{\omega^2}, \tag{13}$$

$$\kappa_1^2 = k_s^2 - k^2\varepsilon_1, \quad \kappa_2^2 = k_s^2 - k^2(1-\eta). \tag{14}$$

Note that the quantity $1-\eta$ exactly coincides with the expression (9) for the Drude-model permittivity $\varepsilon_2(\omega)$. Therefore, we can still use the “phenomenological” relation $\varepsilon_2 = 1-\eta$ everywhere it is suitable and consider this $\varepsilon_2$ as the “permittivity of the film” but remember that it is not an initial presumption but is derived microscopically. In particular, relations (14) appear to be identical with Eq. (5).

Further, for the S mode, the field expressions (3) in the layers 1 ($x > a$) and 3 ($x < -a$) remain the same while the solution of Eqs. (1), (2), (10), (11) for the region $-a < x < a$ gives:

$$H_y = Ae^{-\kappa_1 a} \frac{k}{k_s} \frac{\cosh \kappa_2 x}{\cosh \kappa_2 a} e^{ik_s z} \tag{15}$$

(coincides with the 1st Eq. (4)),

$$E_x = \frac{A}{\varepsilon_2} e^{-\kappa_1 a} \left( \frac{\cosh \kappa_2 x}{\cosh \kappa_2 a} - \eta \frac{\cosh \gamma x}{\cosh \gamma a} \right) e^{ik_s z}, \quad E_z = i \frac{A}{\varepsilon_2} e^{-\kappa_1 a} \left( \frac{\kappa_2}{k_s} \frac{\sinh \kappa_2 x}{\cosh \kappa_2 a} - \eta \frac{k_s}{\gamma} \frac{\sinh \gamma x}{\cosh \gamma a} \right) e^{ik_s z}, \tag{16}$$

$$v_x = i \frac{e}{m\omega} \frac{A}{\varepsilon_2} e^{-\kappa_1 a} \left( \frac{\cosh \kappa_2 x}{\cosh \kappa_2 a} - \frac{\cosh \gamma x}{\cosh \gamma a} \right) e^{ik_s z}, \quad v_z = \frac{e}{m\omega} \frac{A}{\varepsilon_2} e^{-\kappa_1 a} \left( -\frac{\kappa_2}{k_s} \frac{\sinh \kappa_2 x}{\cosh \kappa_2 a} + \frac{k_s}{\gamma} \frac{\sinh \gamma x}{\cosh \gamma a} \right) e^{ik_s z}, \tag{17}$$

$$n = \frac{A}{4\pi e} e^{-\kappa_1 a} \frac{\eta}{\varepsilon_2} \left( \frac{k_s^2}{\gamma} - \gamma \right) \frac{\sinh \gamma x}{\cosh \gamma a} e^{ik_s z}. \tag{18}$$

Again, results for the AS mode can be obtained from (15) – (18) upon replacements (7), (8) and, additionally,

$$\cosh \gamma x \rightleftarrows \sinh \gamma x . \tag{19}$$

Prior to undertake further calculations, let us make some numerical illustrations. For example, in this paper we suppose that the layers 1 and 3 are formed of silica, the film 2 is of the width 0.03 μm and the electrons' motion herewith can be described by the Drude-model approximation for electron gas in silver [21,30,31]. This gives the following input numerical data:

$$\varepsilon_1 = 2.2,\ a = 15\text{ nm},\ \omega_p = 7.73\cdot 10^{15}\text{ s}^{-1},\ v_F = 1.39\cdot 10^{8}\text{ cm/s},\ \beta^2 = 1.18\cdot 10^{16}\text{ cm}^2/\text{s}^2 . \tag{20}$$

The corresponding dispersion curve for the SPP S mode is determined by Eqs. (12) – (14) and visually coincides with the blue curve of Fig. 2 determined via (6) (similarly, for the AS mode, hyperbolic tangents in Eq. (12) should be replaced by cotangents, and the dispersion law is well described by the green curve). The common horizontal asymptote of both curves corresponds to $\omega_c = 4.32\cdot 10^{15}$ s$^{-1}$.

The relatively weak influence of the microscopic corrections, associated with the presence of $\gamma$ in Eqs (12) – (14) and (16) – (18), is explained by the fact that for the whole range of $k_s$, $\gamma$ is approximately two orders higher than other parameters characterizing the SPP-field spatial distribution ($k$, $k_s$, $\kappa_1$, $\kappa_2$, see Fig. 2). For example, upon the conditions of point B marked by the red asterisk on the blue curve in Fig. 2, the main SPP parameters are

$$k = 1.51\cdot 10^{5}\text{ sm}^{-1},\quad k_s = k_{sB} = 8.207\cdot 10^{-5}\text{ sm}^{-1},\quad \omega = 4.52\cdot 10^{15}\text{ s}^{-1},\quad \varepsilon_2 = -1.92;$$

$$\lambda = 2\pi/k = 0.417\ \mu\text{m},\quad \lambda_s = 2\pi/k_s = 0.0766\ \mu\text{m};$$

$$1/\kappa_2 = 11.8\text{ nm},\quad 1/\kappa_1 = 12.7\text{ nm},\quad 1/\gamma = 0.173\text{ nm},\quad \gamma a = 86.7. \tag{21}$$

(these data are taken from the S-mode conditions but the dashed curves in Fig. 2 show that for the AS mode, values of the main parameters are very close to those for the S mode). Under these conditions, the $\gamma$-dependent term in (12) makes a rather minor numerical correction to the results following from the phenomenological equation (6), and the microscopic refinements of Eqs. (12) – (14) and (16) – (18) weakly influence on the SPP properties illustrated in Fig. 2 (and the SPP dynamical characteristics presented below). In essence of the current presentation, we are interested in the approximation

$$\gamma \to \infty \tag{22}$$

(or, equivalently, $\beta \to 0$). However, the $\beta$- (or $\gamma$-) containing terms with seemingly negligible numerical influence cannot be omitted from the beginning because they account for the principally important inhomogeneity of the medium as well as enable to preserve the field continuity and zero normal velocity of electrons at $x = \pm a$. The corresponding contributions differ from zero only in the nearest vicinity of the boundaries so they are called "near-surface" (NS) terms [21].

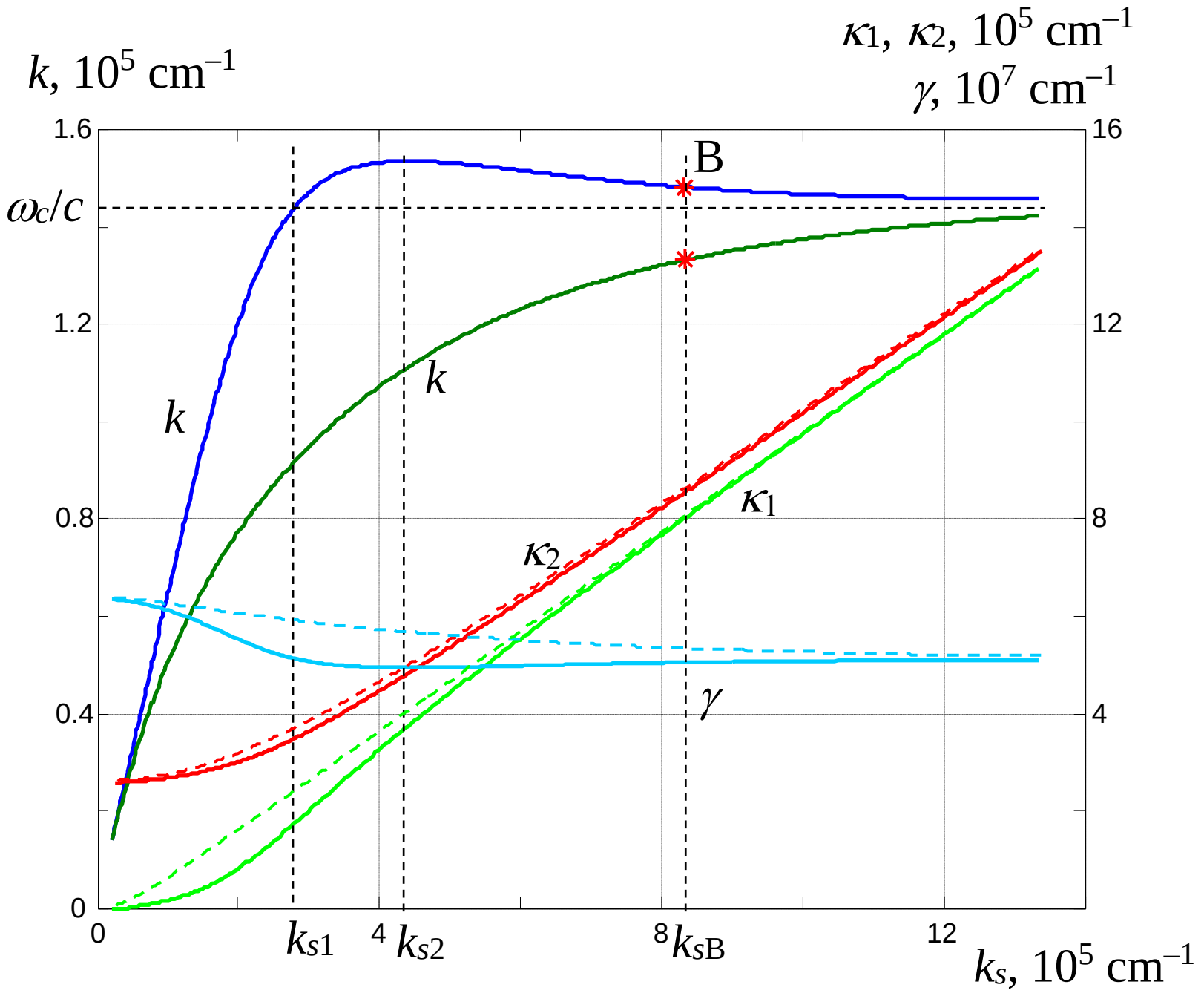

Fig. 2. Wavenumber dependence of the SPP field characteristics calculated via (5), (6) and (12) – (14) for the parameters' values (20): (blue) dispersion curve of the S mode, left vertical scale; (green) dispersion curve of the AS mode, left vertical scale; (light green) $\kappa_1(\omega)$, (red) $\kappa_2(\omega)$, (cyan) $\gamma$ (13): right vertical scale; (solid lines) S mode, (dashed lines) AS mode. Asterisks denote the points B on the dispersion curves for which the field parameters are expressed by (21).

Generally, there are 3 kinds of NS terms [21]. First kind is represented, e.g., by the second summands in parentheses of Eqs. (16), (17) for $E_x$ and $v_x$: they describe rapid changes of the $x$-dependent quantities, but the change "magnitudes" do not depend on $\gamma$. These NS terms characterize physically important details of the spatial variations (e.g., continuity of $E_x$ and vanishing $v_x$ at the boundary) but practically do not affect the "integral" values of the field quantities calculated for the "whole" SPP. In the simplified model of Fig. 1, boundless in the $z$- and $y$-directions, such integral values can only be defined per unit $z$-length and unit $y$-width, and are obtained by the integration over $x$, for example:

$$\langle \ldots \rangle = \int_{-\infty}^{\infty} (\ldots)\, dx\,, \quad \langle \ldots \rangle_2 = \int_{-a}^{a} (\ldots)\, dx\,, \quad \langle \ldots \rangle_{13} = \int_{a}^{\infty} (\ldots)\, dx + \int_{-\infty}^{-a} (\ldots)\, dx \tag{23}$$

(first definition (23) implies the integration over the whole range of $x$ and denotes "total" quantities for the whole SPP field, whereas the second and third ones denote the separate contributions inside the film (medium 2) and in the dielectric (media 1 and 3)). Second kind refers to the "singular" NS terms proportional to $\gamma \cosh(\gamma x)$ or $\gamma \sinh(\gamma x)$ (see Eq. (18)), describing the strictly-localized near-surface contributions whose magnitude grows infinitely upon the condition (22); these terms make meaningful contributions also to the integral values (23). In the approximation (22), due to relations

$$\gamma\frac{\sinh\gamma x}{\cosh\gamma a}\simeq\left[\delta\left(x-a\right)-\delta\left(x+a\right)\right],\quad \gamma\frac{\cosh\gamma x}{\sinh\gamma a}\simeq\left[\delta\left(x-a\right)+\delta\left(x+a\right)\right] \tag{24}$$

(the first combination occurs in expressions for the S mode, see (18), the second one is relevant for the AS mode), they can be reduced to the delta-functions [21]. And finally, the $\gamma^{-1}$-terms, proportional to negative degrees of $\gamma$ (e.g., second summands in expressions for $E_z$ and $v_z$ in Eqs. (16), (17)) are generally very small and can only be held temporarily in some intermediate expressions [21]. These will normally be neglected, and we will call them "insignificant" in contrast to the "significant" NS terms of the two first kinds.

In the present analysis, the NS terms will normally disappear from the final expressions; however, they describe principally important ultra-subwavelength details of the spatial field distributions, and are discussed in the SM (see also [21]).

## 3. Energy and energy flow

Now we proceed to the analysis of the dynamical characteristics: energy, energy flow and momentum of the SPP field. Our study is based on Eqs. (3) – (6) and (12) – (18). For brevity, we explicitly present only the S-mode expressions, which can be transformed for the AS mode via the modifications (8), (19).

Let us start with the energy in dielectric layers whose density $w_{\rm D}$ follows from the standard definition [18,19,21]

$$w_{\rm D}=\frac{1}{16\pi}\left(\varepsilon_1\left|\mathbf{E}\right|^2+\left|\mathbf{H}\right|^2\right).$$

With the help of Eqs. (3) and (23) one finds

$$w_{\rm D}=\frac{g}{\varepsilon_1}e^{-2\kappa_1\left(|x|-a\right)},\quad \left\langle w_{\rm D}\right\rangle=\left\langle w_{\rm D}\right\rangle_{13}=\frac{g}{\varepsilon_1\kappa_1} \tag{25}$$

where the subscript "13" can be omitted because $w_{\rm D}=0$ in the film, and the normalization constant is introduced:

$$g=\frac{\left|A\right|^2}{8\pi}e^{-2\kappa_1 a}=\frac{\left|E_x\left(a\right)\right|^2}{8\pi}. \tag{26}$$

In the metallic layer [23,25]

$$w=\frac{1}{16\pi}\left(\left|\mathbf{E}\right|^2+\left|\mathbf{H}\right|^2\right)+\frac{1}{4}mn_0\left|\mathbf{v}\right|^2+\frac{1}{4}\frac{m\beta^2}{n_0}\left|n\right|^2=w^F+w^M \tag{27}$$

where the first summand expresses the "pure" field contribution $w^F$ whereas the terms containing the electrons' velocity and density express the "material" contribution $w^M$. By using Eqs. (15) – (18) and neglecting the non-singular NS terms, we obtain

$$w^F=\frac{g}{2}\frac{1}{\varepsilon_2^2}\left[\left(1+\varepsilon_2^2\frac{k^2}{k_s^2}\right)\frac{\cosh^2\kappa_2 x}{\cosh^2\kappa_2 a}+\frac{\kappa_2^2}{k_s^2}\frac{\sinh^2\kappa_2 x}{\cosh^2\kappa_2 a}\right], \tag{28}$$

$$w^M=\frac{g}{2}\frac{\eta}{\varepsilon_2^2}\left[\frac{\cosh^2\kappa_2 x}{\cosh^2\kappa_2 a}+\frac{\kappa_2^2}{k_s^2}\frac{\sinh^2\kappa_2 x}{\cosh^2\kappa_2 a}\right]. \tag{29}$$

The integral values are then easily derived according to (23):

$$\left\langle w^F\right\rangle_2=-g\frac{1}{2\varepsilon_2}\left[\frac{\kappa_1}{\kappa_2^2\varepsilon_1}\left(2+\frac{k^2}{k_s^2}\varepsilon_2\left(\varepsilon_2-1\right)\right)-\left(\varepsilon_2+1\right)\frac{k^2}{k_s^2}\frac{a}{\cosh^2\kappa_2 a}\right], \tag{30}$$

$$\left\langle w^M\right\rangle_2=-g\frac{\eta}{2\varepsilon_2}\left[\frac{\kappa_1}{\kappa_2^2\varepsilon_1}\left(2-\frac{k^2}{k_s^2}\varepsilon_2\right)-\frac{k^2}{k_s^2}\frac{a}{\cosh^2\kappa_2 a}\right]. \tag{31}$$

Therefore, the "full" energy in the metal film $\langle w\rangle_2 = \langle w^F\rangle_2 + \langle w^M\rangle_2$ is

$$\langle w\rangle_2 = g\left[\frac{\kappa_1}{\kappa_2^2\varepsilon_1}\left(-\frac{1+\eta}{\varepsilon_2}+\eta\frac{k^2}{k_s^2}\right)+\frac{k^2}{\varepsilon_2 k_s^2}\frac{a}{\cosh^2\kappa_2 a}\right],$$

and whole integral SPP energy follows hence after adding (25),

$$\langle w\rangle = g\left[\frac{1}{\kappa_1\varepsilon_1}+\frac{\kappa_1}{\kappa_2^2\varepsilon_1}\left(-\frac{1+\eta}{\varepsilon_2}+\eta\frac{k^2}{k_s^2}\right)+\frac{k^2}{\varepsilon_2 k_s^2}\frac{a}{\cosh^2\kappa_2 a}\right]. \tag{32}$$

For the AS mode, expression (32) modifies to the form

$$\langle w\rangle \to \langle w_{AS}\rangle = g\left[\frac{1}{\kappa_1\varepsilon_1}+\frac{\kappa_1}{\kappa_2^2\varepsilon_1}\left(-\frac{1+\eta}{\varepsilon_2}+\eta\frac{k^2}{k_s^2}\right)-\frac{k^2}{\varepsilon_2 k_s^2}\frac{a}{\sinh^2\kappa_2 a}\right] \tag{33}$$

which illustrates the general rule: The "integral" values (23) for the AS mode can be obtained from the explicitly presented S-mode expressions where, in addition to substitutions (8), the replacement $a \to -a$ is performed. The behavior of the energy contributions (25), (30), (31) and (32) is illustrated by the green, red, cyan and magenta curves in Figs. 3a, b.

The energy flow density is determined by the usual Poynting vector expression [18,19,21]

$$\mathbf{S} = \frac{c}{8\pi}\mathrm{Re}\left(\mathbf{E}^* \times \mathbf{H}\right) = \mathbf{z}\frac{c}{8\pi}E_x^* H_y \tag{34}$$

which gives, in the dielectric layers,

$$S_\mathrm{D} = c\frac{g}{\varepsilon_1}\frac{k}{k_s}e^{-2\kappa_1(|x|-a)} \tag{35}$$

and in the film

$$S = c\frac{g}{\varepsilon_2}\frac{k}{k_s}\frac{\cosh^2\kappa_2 x}{\cosh^2\kappa_2 a}. \tag{36}$$

Note that the energy flow in dielectric (35) is directed along the wave propagation whereas in the film volume, $S$ (36) is always directed against the propagation (but changes the sign together with $E_x$ in the closest vicinity of the boundary due to the NS term which is not shown here, see SM).

Eqs. (35) and (36) directly lead to the known expression for integral energy flow [28,29]

$$\langle S\rangle = cg\frac{k}{k_s}\left(\frac{1}{\varepsilon_2}\frac{a}{\cosh^2\kappa_2 a}-\frac{\kappa_1}{\kappa_2^2\varepsilon_1}+\frac{1}{\varepsilon_1\kappa_1}\right) \tag{37}$$

(blue curves in Figs. 3a, b). The above formulas show that the total energy flow is formed as a combination of oppositely directed contributions. In the long-wavelength region (while $k_s < k_{s2}$, cf. Fig. 2), the positive flow in dielectric layers prevails and the whole integral flow agrees with the wave propagation but when $k_s > k_{s2}$ the regime of "back flow" is realized for the S mode [28,29]. This is the most interesting feature of the S-mode SPP (Fig. 3a).

Based on the total energy (32) and the energy flow (37), the group velocity of the SPP can be found in the form

$$v_g = \frac{\langle S\rangle}{\langle w\rangle} = c\frac{k}{k_s}\frac{1-\dfrac{\kappa_1^2}{\kappa_2^2}+\dfrac{\varepsilon_1}{\varepsilon_2}\dfrac{\kappa_1 a}{\cosh^2\kappa_2 a}}{1+\dfrac{\kappa_1^2}{\kappa_2^2}\left(-\dfrac{1+\eta}{\varepsilon_2}+\eta\dfrac{k^2}{k_s^2}\right)+\dfrac{\varepsilon_1 k^2}{\varepsilon_2 k_s^2}\dfrac{\kappa_1 a}{\cosh^2\kappa_2 a}} \tag{38}$$

which is illustrated by the brown curves in Figs. 3a, b. This result expectedly agrees with what follows from the formal group-velocity definition [18,19] $v_g = d\omega/dk_s$.

The blue and brown curves in Fig. 3a spectacularly show the sign reversal of the energy flow (37) and group velocity (38) of the S mode in the point $k_{s2}$ where the blue dispersion curve of Fig. 2 has the maximum. In contrast, for the AS mode, the energy flow and group velocity are always positive. Other remarkable differences between Figs. 3a and 3b are the much smaller value of $\langle w^F \rangle_2$ in case of the AS mode (red curves) and the tendency of $\langle w^F \rangle_2$ in the S mode to zero with decreasing frequency (cyan curve in Fig. 3a).

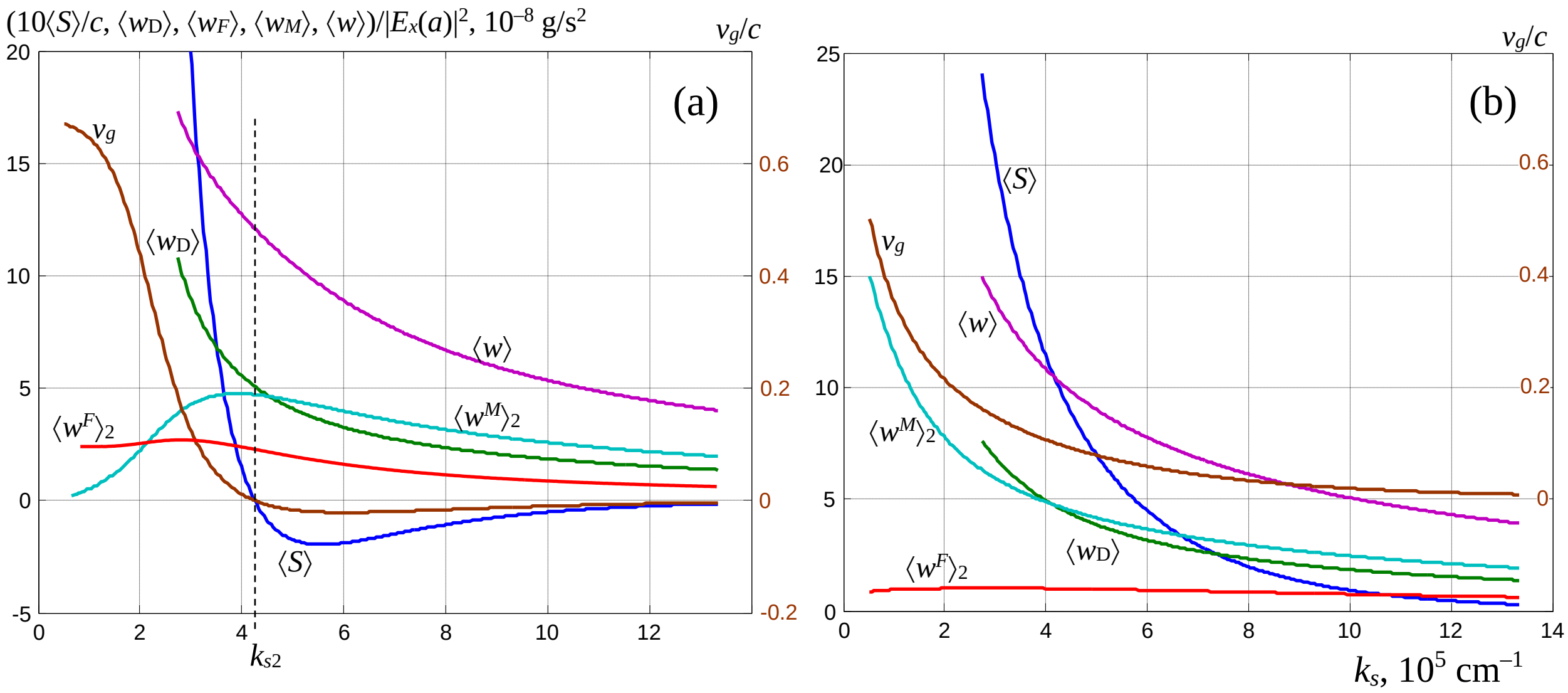

Fig. 3. Wavenumber dependences of the (blue) integral energy flow (37) and the integral energy contributions: (green) in dielectric (25), (red) “field” contribution in metal (30), (cyan) material contribution in metal (31), (magenta) total energy (24), and (brown) the group velocity (38) (right scale); (a) for the S mode, (b) for the AS mode. Other accepted conditions are as in Fig. 2.

Curves in Fig. 3 represent the energy and energy flow constituents normalized by $|E_x(a)|^2$, which makes an impression that the quantities decrease with growing $k_s$. However, $|E_x(a)|^2$ is not fixed, and the behavior presented in Fig. 3 testifies mainly for the growth of the electric (and magnetic) field values near the interface due to the growing field confinement at high $k_s$ [2,6].

### 4. Momentum of the propagating mode

The momentum components, as the vector quantities, are denoted by the bold “**p**” furnished with relevant sub- and superscripts; simultaneously, as all of them are collinear to the *z*-axis, here and further the “bold” vector notation will be frequently replaced by the corresponding scalar “*p*” with the same sub- and superscripts, following the simple rule

$$\mathbf{p}_{\oplus}^{\oplus} = \mathbf{z} p_{\oplus}^{\oplus}$$

where “$\oplus$” means arbitrary combination of symbols and **z** is the unit vector in the *z*-direction. As in the Section 3, we explicitly present only the S-mode expressions, keeping in mind that their AS-mode counterparts can be formally obtained via replacements (8) and $a \to -a$, according to the remark beneath Eq. (33).

Also, as a general rule, we emphasize that in the symmetric structure of Fig. 1, all the momentum constituents are distributed symmetrically with respect to the middle plane $x = 0$, both in the S and AS modes.

4.1. In dielectric layers $|x| > a$ the momentum is determined routinely as the Minkowski momentum [15,20,21] (see Eqs. (34), (35))

$$p_{\mathrm{D}} = \frac{\varepsilon_1}{c^2} S_{\mathrm{D}} = \frac{g}{c} \frac{k}{k_s} e^{-2\kappa_1(|x|-a)} \quad . \tag{39}$$

It can be divided into the orbital and spin contributions [20,21,27]

$$p_{\mathrm{DO}} = \frac{g}{\omega} \frac{k_s}{\varepsilon_1} e^{-2\kappa_1(|x|-a)}, \quad p_{\mathrm{D}S} = -\frac{g}{\omega} \frac{\kappa_1^2}{k_s \varepsilon_1} e^{-2\kappa_1(|x|-a)}. \tag{40}$$

For the integral values (23), by using the simplified notation according to the note below Eq. (25), we obtain

$$\langle p_{\mathrm{DO}} \rangle = \frac{g}{\omega} \frac{k_s}{\kappa_1 \varepsilon_1}, \quad \langle p_{\mathrm{D}S} \rangle = -\frac{g}{\omega} \frac{\kappa_1}{k_s \varepsilon_1}, \quad \langle p_{\mathrm{D}} \rangle = \frac{g}{\omega} \frac{k^2}{k_s \kappa_1}. \tag{41}$$

Typically for the evanescent waves, the spin and orbital contributions are directed oppositely [7,8,15,20]. The behavior of the momentum constituents can be suitably analyzed in the normalized units $c\langle p \rangle / \langle w \rangle$ ("per plasmon" [15,16]), and this presentation will be systematically employed below. For the momentum constituents in dielectric (41), this behavior is illustrated in Figs. 4a, b for the S and AS modes, correspondingly. Note that in the low-frequency limit $c\langle p_{\mathrm{D}} \rangle \to \langle w \rangle \sqrt{\varepsilon_1}$ in agreement with the Minkowski picture.

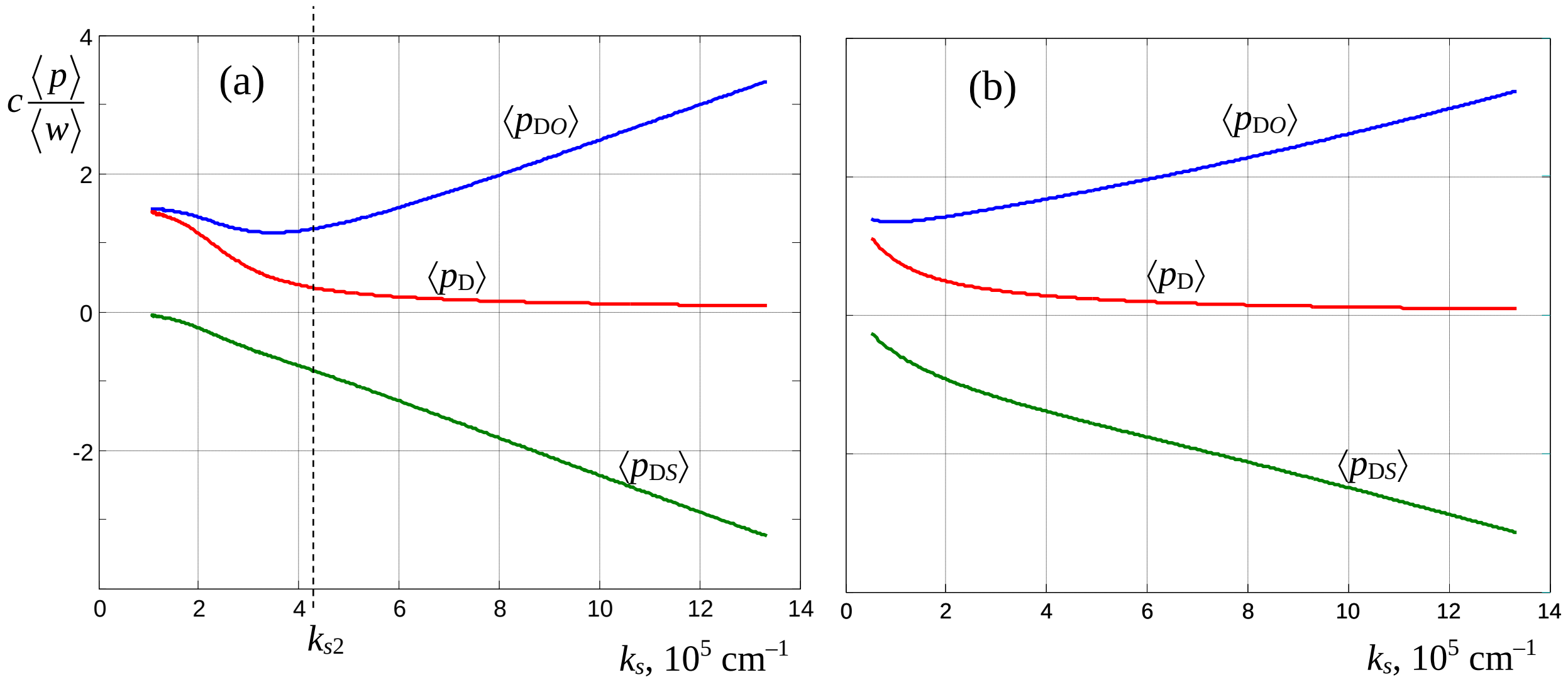

Fig. 4. Orbital (canonical), spin and total momentum in dielectric (outside the metal film): (a) S mode, (b) AS mode.

In Fig. 4, the differences between the S mode and the AS mode is well noticeable in the region $k_s < k_{s2}$ where the two dispersion curves differ essentially (cf. Fig. 2). For short SPP waves (high $k_s$), the momentum components of the S and AS modes show rather similar performance.

4.2. Momentum in the metal film. According to [15,21], the momentum in metal $\mathbf{p}$ consists of several contributions which can be classified in different ways. We start with separation of the "field" (electromagnetic) and "material" parts: $\mathbf{p} = \mathbf{p}^F + \mathbf{p}^M$ [21]. The "pure field" momentum is described by the Poynting vector (34)

$$p^F = \frac{1}{c^2} S,$$

which for the field (3), (15) – (18) results in

$$p^F = \frac{1}{8\pi c} E_x^* H_y = \frac{g}{c} \frac{1}{\varepsilon_2} \frac{k}{k_s} \frac{\cosh^2 \kappa_2 x}{\cosh^2 \kappa_2 a}, \quad \left\langle p^F \right\rangle_2 = \frac{g}{\omega} \frac{1}{\varepsilon_2} \frac{k^2}{k_s \kappa_2} \left( \frac{\kappa_2 a}{\cosh^2 \kappa_2 a} - \frac{\kappa_1 \varepsilon_2}{\kappa_2 \varepsilon_1} \right) \tag{42}$$

The frequency-dependent behavior of the "field" momentum (42) under conditions accepted for Fig. 2 is illustrated in Fig. 5a, b (red curves). In both S and AS modes the "field" momentum contribution is directed negatively because of the negative $\varepsilon_2$ but its absolute value is comparatively small due to the multiplier $k^2/k_s\kappa_2$. For the S mode, $\langle p^F \rangle_2$ acquires relatively large values in the positive-flow region ($k_s < k_{s2}$) and its absolute value is generally higher than in the AS mode.

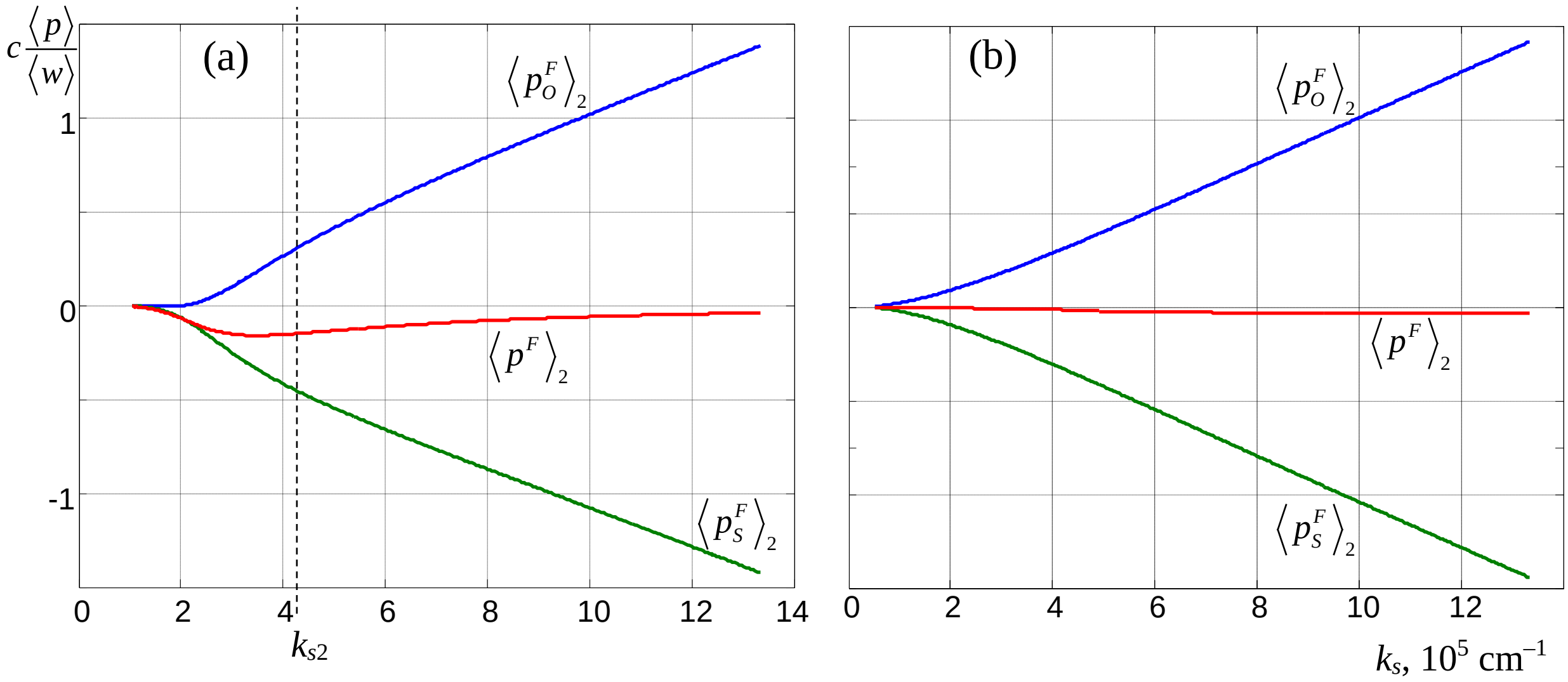

Fig. 5. "Field" momentum (42) and its orbital and spin constituents (60), (61) (a) in the S mode and (b) in the AS mode under conditions of Fig. 2.

Material contribution to the momentum is $\mathbf{p}^M = \mathbf{p}_m^M + \mathbf{p}_e^M + \mathbf{p}_{SR}^M$ [21] where $\mathbf{p}_m^M$ and $\mathbf{p}_e^M$ originate from the linear motion of electrons. The first constituent appears due to interaction of the field-induced dipole moments with the magnetic field of the wave [15,21]:

$$\mathbf{p}_m^M = \frac{1}{2c}\left(\mathbf{d}^* \times \mathbf{H}\right) \Rightarrow \frac{\alpha}{2c} \mathrm{Re}\left(\mathbf{E}^* \times \mathbf{H}\right) = \frac{\varepsilon_2 - 1}{8\pi c} \mathrm{Re}\left(\mathbf{E}^* \times \mathbf{H}\right), \tag{43}$$

$$p_m^M = \frac{\alpha}{2c} E_x^* H_y = -\frac{g}{c} \frac{\eta}{\varepsilon_2} \frac{k}{k_s} \frac{\cosh^2 \kappa_2 x}{\cosh^2 \kappa_2 a}, \quad \left\langle p_m^M \right\rangle = -\frac{g}{\omega} \frac{\eta}{\varepsilon_2} \frac{k^2}{k_s \kappa_2} \left( \frac{\kappa_2 a}{\cosh^2 \kappa_2 a} - \frac{\kappa_1 \varepsilon_2}{\kappa_2 \varepsilon_1} \right) \tag{44}$$

($\alpha = (\varepsilon_2 - 1)/4\pi = -\eta/4\pi$ is the polarizability which expresses proportionality between the dipole momentum density $\mathbf{d} = n_0 \left(ie/\omega\right) \mathbf{v}$ and the electric field, $\mathbf{d} = \alpha \mathbf{E}$). Actually, $\alpha$ is a function of $x$ and may be singular (see [21] and the SM), but the equality (43) neglects this circumstance; the symbol "$\Rightarrow$" implies the "coarse" correspondence, neglecting the NS terms inessential for the integral values. Like in Eqs. (41), we use the abbreviated notation, $\langle \ldots \rangle_2 \to \langle \ldots \rangle$, because the material terms do not exist in the dielectric layers. The momentum component (44) is illustrated by the blue curves in Fig. 6a, b. It is rather remarkable for the S mode in the positive-flow region ($k_s < k_{s2}$) but in the AS mode this contribution is practically the smallest compared to other material contributions (cf. Fig. 6b).

The next material part of the momentum appears due to interactions between the induced dipoles and the electric vector of the wave. It can be represented in the form [15]

$$\mathbf{p}_e^M = \mathbf{p}_{ex}^M + \mathbf{p}_{ez}^M \Rightarrow \frac{1}{16\pi}\frac{d\varepsilon_2}{d\omega}\mathrm{Im}\left[\mathbf{E}^*\cdot(\nabla)\mathbf{E}\right] \tag{45}$$

(cf. Eq. (S29) in the SM) which, again, follows from the exact relations [21] after omitting the NS terms. For the SPP field (3) – (6), this equation gives

$$p_{ex}^M = \frac{k_s}{4}\frac{d\alpha}{d\omega}\left|E_x\right|^2 = \frac{g}{c}\frac{\eta}{\varepsilon_2^2}\frac{k_s}{k}\frac{\cosh^2\kappa_2 x}{\cosh^2\kappa_2 a}, \quad \left\langle p_{ex}^M\right\rangle = \frac{g}{\omega}\frac{\eta}{\varepsilon_2^2}\frac{k_s}{\kappa_2}\left(\frac{\kappa_2 a}{\cosh^2\kappa_2 a} - \frac{\kappa_1\varepsilon_2}{\kappa_2\varepsilon_1}\right); \tag{46}$$

$$p_{ez}^M = \frac{k_s}{4}\frac{d\alpha}{d\omega}\left|E_z\right|^2 = \frac{g}{c}\frac{\eta}{\varepsilon_2^2}\frac{\kappa_2^2}{kk_s}\frac{\sinh^2\kappa_2 x}{\cosh^2\kappa_2 a}, \quad \left\langle p_{ez}^M\right\rangle = -\frac{g}{\omega}\frac{\eta}{\varepsilon_2^2}\frac{\kappa_2}{k_s}\left(\frac{\kappa_2 a}{\cosh^2\kappa_2 a} + \frac{\kappa_1\varepsilon_2}{\kappa_2\varepsilon_1}\right). \tag{47}$$

Notably, for the S mode $\left\langle p_{ex}^M\right\rangle > \left\langle p_{ez}^M\right\rangle$ whereas for the AS mode, to the opposite, $\left\langle p_{ex}^M\right\rangle < \left\langle p_{ez}^M\right\rangle$ in the whole range of the wavelengths (see the red and green curves in Figs. 6a and 6b). This is because in the S mode $E_z(x)$ is an odd function and therefore close to zero in the film depth ($x \approx 0$), so its contribution to the integral value is smaller than that of the even function $E_x(x)$, which determines $\left\langle p_{ex}^M\right\rangle$. For the AS modes, just on the contrary, $E_z(x)$ is even and $E_x(x)$ is odd. In the sum

$$\left\langle p_e^M\right\rangle = \left\langle p_{ex}^M\right\rangle + \left\langle p_{ez}^M\right\rangle = \frac{g}{\omega}\frac{\eta}{\varepsilon_2^2}\frac{k_s}{\kappa_2}\left[\frac{k^2}{k_s^2}\frac{\kappa_2 a}{\cosh^2\kappa_2 a} - \frac{\kappa_1}{\kappa_2\varepsilon_1}\left(2-\varepsilon_2\frac{k^2}{k_s^2}\right)\right], \tag{48}$$

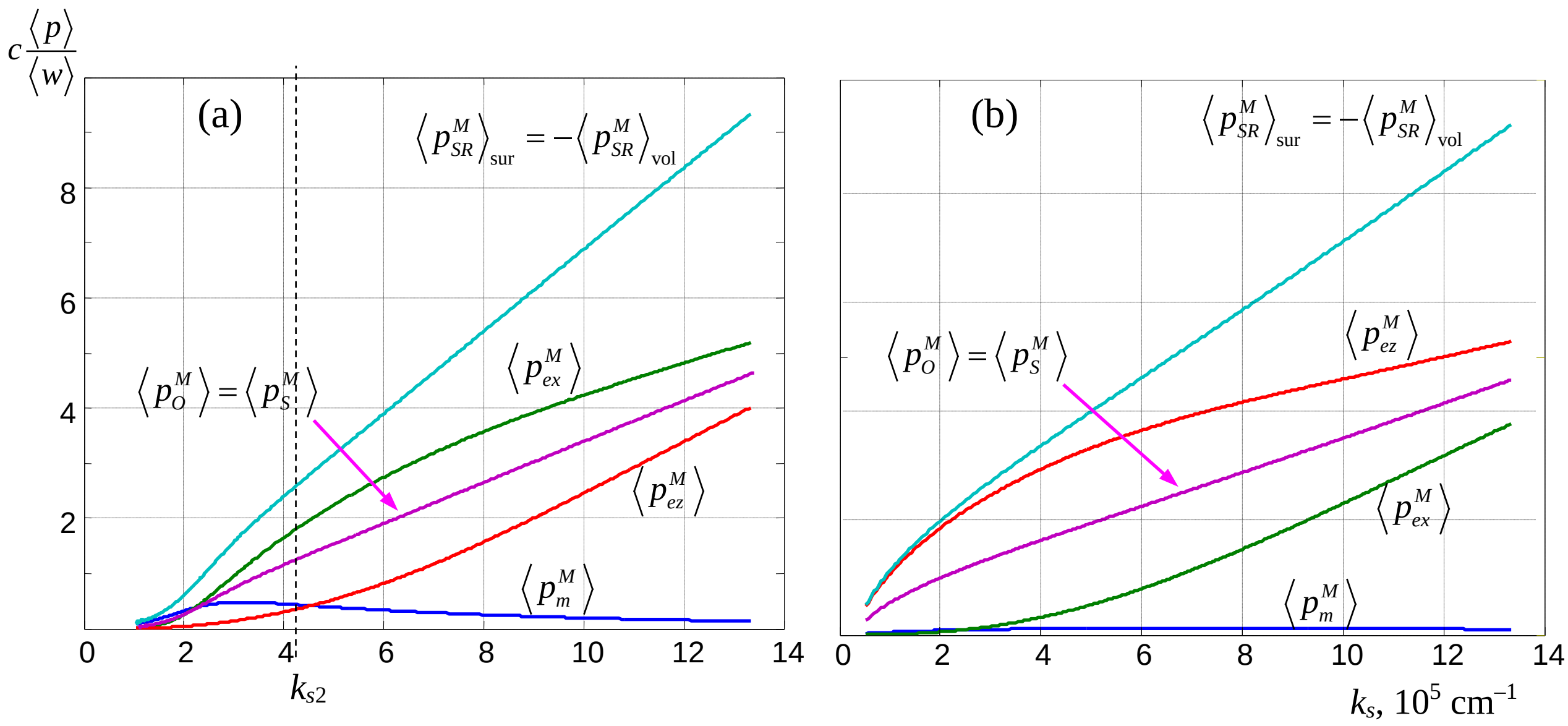

Fig. 6. Separate material contributions of the momentum in the metallic film $|x| < a$ (a) for the S mode and (b) for the AS mode: (blue) $\left\langle p_m^M\right\rangle$ (44), (green) $\left\langle p_{ex}^M\right\rangle$ (46), (red) $\left\langle p_{ez}^M\right\rangle$ (47), (cyan) $\left\langle p_{SR}^M\right\rangle_{\mathrm{sur}} = -\left\langle p_{SR}^M\right\rangle_{\mathrm{vol}}$ (51), (52), (magenta) spin and orbital material momenta $\left\langle p_O^M\right\rangle = \left\langle p_S^M\right\rangle$ (63), (64).

these discrepancies partly compensate each other so that only the small difference remains between the S-mode and the AS-mode behaviors. It is associated with the first term in brackets of (48), which tends to zero with growing $k_s$.

Finally, $\mathbf{p}_{SR}^M$ is the contribution of the rotational motion of electrons in the metal [15,21], which upon neglecting the NS terms can be found as

$$\mathbf{p}_{SR}^{M} = \frac{1}{2}\left[\frac{mn_0}{2\omega}\nabla\times\mathrm{Im}\left(\mathbf{v}^{*}\times\mathbf{v}\right)\right] \Rightarrow \frac{1}{16\pi}\frac{d\varepsilon_2}{d\omega}\nabla\times\mathrm{Im}\left(\mathbf{E}^{*}\times\mathbf{E}\right) \tag{49}$$

(this momentum emerges as $\mathbf{p}_{SR}^{M} = \frac{1}{2}\nabla\times\mathbf{s}_{R}^{M} = \mathbf{z}\frac{1}{2}\frac{ds_{Ry}^{M}}{dx}$ where $\mathbf{s}_{R}^{M} = \frac{mn_0}{2\omega}\mathrm{Im}\left(\mathbf{v}^{*}\times\mathbf{v}\right)$ is the material spin associated with the electrons' rotation [15,21]). The momentum (49) contains the singular part strictly localized at the surface (owing to the phenomenological spin discontinuity at $x = \pm a$ [17,20]) and the volume part,

$$\mathbf{p}_{SR}^{M} = \left(\mathbf{p}_{SR}^{M}\right)_{\mathrm{sur}} + \left(\mathbf{p}_{SR}^{M}\right)_{\mathrm{vol}}. \tag{50}$$

Consistent calculation of this momentum employs the microscopic equations (16) – (18) (see SM, Eq. (S25)), and with using Eq. (24) one finds

$$\left(p_{SR}^{M}\right)_{\mathrm{sur}} = -\frac{g}{\omega}\frac{\eta}{\varepsilon_2\varepsilon_1}\frac{\kappa_1}{k_s}\left[\delta\left(x-a\right)+\delta\left(x+a\right)\right], \quad \left\langle p_{SR}^{M}\right\rangle_{\mathrm{sur}} = -\frac{2g}{\omega}\frac{\eta}{\varepsilon_2\varepsilon_1}\frac{\kappa_1}{k_s}; \tag{51}$$

$$\left(p_{SR}^{M}\right)_{\mathrm{vol}} = -\frac{g}{\omega}\frac{\eta}{\varepsilon_2^2}\frac{\kappa_2^2}{k_s}\frac{\cosh^2\kappa_2 x+\sinh^2\kappa_2 x}{\cosh^2\kappa_2 a}, \quad \left\langle p_{SR}^{M}\right\rangle_{\mathrm{vol}} = \frac{2g}{\omega}\frac{\eta}{\varepsilon_2\varepsilon_1}\frac{\kappa_1}{k_s}. \tag{52}$$

Remarkably, the surface and volume parts of the material spin momentum exactly compensate each other. This is because both momentum contributions express bound currents formed by the cyclic electron trajectories (analogs of the "Ampere loops" [32,33]), and the "net" transfer of mass by the whole motion is absent [15,17]. The wavenumber dependence of (51), (52) is illustrated by the cyan curves in Figs. 6a, b.

Now we are in a position to establish another remarkable fact, following from the above analysis of the material momentum contributions (44), (46), (47), (50) – (52): the total volume part of the material momentum contribution is zero:

$$\left(p^{M}\right)_{\mathrm{vol}} = p_{m}^{M} + p_{ex}^{M} + p_{ez}^{M} + \left(p_{SR}^{M}\right)_{\mathrm{vol}} = 0. \tag{53}$$

This means that the whole material momentum inside the metallic film reduces to the surface singular term (51),

$$p^{M} = -\frac{g}{\omega}\frac{\eta}{\varepsilon_2\varepsilon_1}\frac{\kappa_1}{k_s}\left[\delta\left(x-a\right)+\delta\left(x+a\right)\right], \quad \left\langle p^{M}\right\rangle = -\frac{2g}{\omega}\frac{\eta}{\varepsilon_2\varepsilon_1}\frac{\kappa_1}{k_s}, \tag{54}$$

and is directed along the wave propagation. Note that the properties (53), (54) of the 3-layer structure are in full correspondence with the analogous properties of the momentum in the usual single-interface SPP [21].

4.3. Spin-orbital momentum decomposition in the film. Another useful classification employs the spin-orbital (canonical) decomposition [15,27]. For the "pure-field" momentum of (42) the canonical decomposition in presence of charges and currents reads $\mathbf{p}^{F} = \mathbf{p}_{S}^{F} + \mathbf{p}_{O}^{F}$ where [15,21]

$$\mathbf{p}_{S}^{F} = \frac{1}{32\pi\omega}\mathrm{Im}\left[\nabla\times\left(\mathbf{E}^{*}\times\mathbf{E}\right)+\nabla\times\left(\mathbf{H}^{*}\times\mathbf{H}\right)\right] - \frac{e}{4\omega}\mathrm{Im}\left(\mathbf{E}^{*}n\right), \tag{55}$$

$$\mathbf{p}_{O}^{F} = \frac{1}{16\pi\omega}\mathrm{Im}\left[\mathbf{E}^{*}\cdot\left(\nabla\right)\mathbf{E}+\mathbf{H}^{*}\cdot\left(\nabla\right)\mathbf{H}\right] - \frac{n_0 e}{4\omega c}\mathrm{Im}\left(\mathbf{H}^{*}\times\mathbf{v}\right). \tag{56}$$

For the considered TM waves, the term with $\mathbf{H}^{*}\times\mathbf{H}$ vanishes, and the singular NS terms in $\mathrm{Im}\left(\mathbf{E}^{*}n\right)$ and $\mathrm{Im}\left[\nabla\times\left(\mathbf{E}^{*}\times\mathbf{E}\right)\right]$ of Eq. (55) mutually cancel [21], so the final expression of the "field" spin momentum, with omitted NS terms is

$$\mathbf{p}_{S}^{F} = \frac{1}{32\pi\omega}\mathrm{Im}\left[\nabla\times\left(\mathbf{E}^{*}\times\mathbf{E}\right)\right]. \tag{57}$$

In Eq. (56), the term with $\mathrm{Im}\left(\mathbf{H}^{*}\times\mathbf{v}\right)$ can be transformed because for the TM modes

$$\frac{1}{8\pi c}\mathrm{Re}\left[\mathbf{E}^{*}\times\mathbf{H}\right]=\frac{1}{16\pi\omega}\mathrm{Im}\left[\nabla\times\left(\mathbf{H}^{*}\times\mathbf{H}\right)\right]+\frac{1}{8\pi\omega}\mathrm{Im}\left[\mathbf{H}^{*}\cdot(\nabla)\mathbf{H}\right]-\frac{n_0 e}{2\omega c}\mathrm{Im}\left(\mathbf{H}^{*}\times\mathbf{v}\right)$$

$$=\frac{1}{8\pi\omega}\mathrm{Im}\left[\mathbf{H}^{*}\cdot(\nabla)\mathbf{H}\right]-\frac{n_0 e}{2\omega c}\mathrm{Im}\left(\mathbf{H}^{*}\times\mathbf{v}\right);$$

$$\frac{1}{8\pi c}\mathrm{Re}\left[\mathbf{E}^{*}\times\mathbf{H}\right]=-\frac{1}{8\pi c}\mathrm{Re}\left[-i\frac{e}{\omega}\alpha^{-1}\mathbf{v}^{*}\times\mathbf{H}\right]=-\frac{m\omega}{8\pi e c}\mathrm{Im}\left[\mathbf{H}^{*}\times\mathbf{v}\right],$$

and

$$-\frac{n_0 e}{4\omega c}\mathrm{Im}\left(\mathbf{H}^{*}\times\mathbf{v}\right)=\frac{1}{16\pi\omega}\frac{\eta}{\varepsilon_2}\mathrm{Im}\left[\mathbf{H}^{*}\cdot(\nabla)\mathbf{H}\right]. \tag{58}$$

As a result, the “pure field” orbital momentum (56) acquires the form

$$\mathbf{p}_O^F=\frac{1}{16\pi\omega}\mathrm{Im}\left[\mathbf{E}^{*}\cdot(\nabla)\mathbf{E}+\frac{1}{\varepsilon_2}\mathbf{H}^{*}\cdot(\nabla)\mathbf{H}\right]. \tag{59}$$

Finally, by using Eqs. (4) – (6), the explicit expressions are derived:

$$p_S^F=-\frac{g}{2\omega}\frac{\kappa_2^2}{\varepsilon_2^2 k_s}\frac{\cosh^2\kappa_2 x+\sinh^2\kappa_2 x}{\cosh^2\kappa_2 a},\quad \left\langle p_S^F\right\rangle_2=\frac{g}{\omega}\frac{\kappa_1}{\varepsilon_2\varepsilon_1 k_s}; \tag{60}$$

$$p_O^F=\frac{g}{2\omega}\frac{k_s}{\varepsilon_2^2}\left[\left(1+\varepsilon_2\frac{k^2}{k_s^2}\right)\frac{\cosh^2\kappa_2 x}{\cosh^2\kappa_2 a}+\frac{\kappa_2^2}{k_s^2}\frac{\sinh^2\kappa_2 x}{\cosh^2\kappa_2 a}\right],$$

$$\left\langle p_O^F\right\rangle_2=\frac{g}{\omega}\frac{k^2}{\varepsilon_2 k_s}\left(\frac{a}{\cosh^2\kappa_2 a}-\frac{\kappa_1\varepsilon_2}{\kappa_2^2\varepsilon_1}-\frac{\kappa_1}{k^2\varepsilon_1}\right). \tag{61}$$

Application of the same Eqs. (58) and (59) to (43), (44) supplies the spin-orbital decomposition of the material momentum contribution $p_m^M=\left(\varepsilon_2-1\right)p^F$:

$$p_{mO}^M=-\eta p_O^F,\quad p_{mS}^M=-\eta p_S^F. \tag{62}$$

(see the note beneath Eq. (14)). However, the full material orbital momentum $p_O^M$ additionally includes the contribution $p_{ex}^M+p_{ez}^M$ (45), (48) [21], which gives

$$p_O^M=p_{mO}^M+p_{ex}^M+p_{ey}^M=\frac{g}{2\omega}\frac{\eta}{\varepsilon_2^2}\frac{\kappa_2^2}{k_s}\frac{\cosh^2\kappa_2 x+\sinh^2\kappa_2 x}{\cosh^2\kappa_2 a},\quad \left\langle p_O^M\right\rangle=-\frac{g}{\omega}\frac{\eta}{\varepsilon_2\varepsilon_1}\frac{\kappa_1}{k_s}. \tag{63}$$

On the other hand, the momentum from the rotational motion of electrons (50) should be added to the spin momentum [21], whence the full material spin momentum is

$$p_S^M=p_{mS}^M+p_{SR}^M=-\frac{g}{2\omega}\eta\frac{1}{\varepsilon_2^2}\left(\frac{\kappa_2^2}{k_s}\frac{\cosh^2\kappa_2 x+\sinh^2\kappa_2 x}{\cosh^2\kappa_2 a}+\frac{2\kappa_1\varepsilon_2}{k_s\varepsilon_1}\left[\delta(x-a)+\delta(x+a)\right]\right),$$

$$\left\langle p_S^M\right\rangle=-\frac{g}{\omega}\frac{\eta}{\varepsilon_2\varepsilon_1}\frac{\kappa_1}{k_s}. \tag{64}$$

Interestingly, comparison of Eqs. (63) and (64) shows that the integral material contributions to the orbital and spin momenta are equal. Moreover, the material spin momentum (64) consists of the volume part $p_{mS}^M+\left(p_{SR}^M\right)_{\mathrm{vol}}$ (see (52)), which exactly compensates the orbital contribution (63), and we again arrive at the conclusion that it is the surface part $\left(p_{SR}^M\right)_{\mathrm{sur}}$ (51) that remains as the total material momentum (54).

The spin and orbital contributions (49), (50) of the material momentum are illustrated by the magenta lines in Fig. 6a, b. They indicate no essential differences between the S- and AS-mode behaviors.

Finally, the total orbital momentum in the film is determined by the sum of the "field" (61) and material (63) contributions, $p_O = p_O^F + p_O^M$,

$$\langle p_O \rangle_2 = \langle p_O^F \rangle_2 + \langle p_O^M \rangle = \frac{g}{\omega} \frac{k^2}{\varepsilon_2 k_s} \left( \frac{a}{\cosh^2 \kappa_2 a} - \frac{\kappa_1 \varepsilon_2}{\kappa_2^2 \varepsilon_1} - 2\frac{\kappa_1}{k^2 \varepsilon_1} + \frac{\kappa_1}{k^2} \frac{\varepsilon_2}{\varepsilon_1} \right); \tag{65}$$

likewise, the total spin momentum in the layer 2 $p_S = p_S^M + p_S^F$ is formed by (64) and (60), which results in

$$\langle p_S \rangle_2 = \langle p_S^F \rangle_2 + \langle p_S^M \rangle = \frac{g}{\omega} \frac{\kappa_1}{\varepsilon_1 k_s}. \tag{66}$$

The behavior of the momentum "blocks" (65) and (66) is illustrated by the magenta and brown curves in Fig. 7a, b.

4.4. Spin-orbital momentum constituents in the whole structure. Now we summarize the data on the spin-orbital momentum decomposition for the whole SPP wave supported by the symmetric 3-layer structure of Fig. 1. The main results are presented in Fig. 7, together with the orbital and spin momenta in the dielectric $\langle p_{DO} \rangle$, $\langle p_{DS} \rangle$ (see also Fig. 4a, b) and in the metal film $\langle p_O \rangle_2$, $\langle p_S \rangle_2$.

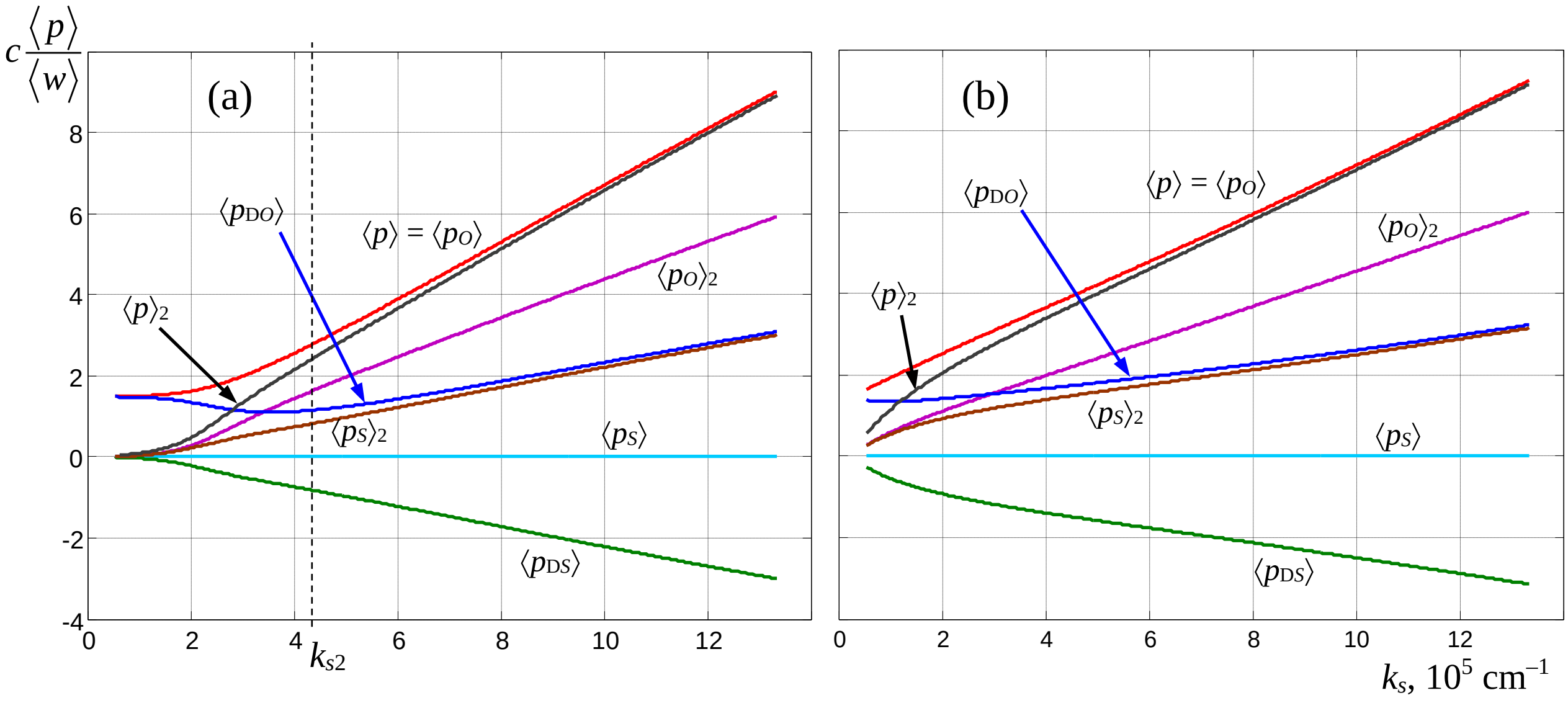

Fig. 7. Orbital and spin momentum constituents of the SPP wave for (a) the S mode and (b) AS mode: (blue) orbital momentum in dielectric $|x| > a$, (green) spin momentum in dielectric (blue and green curves are reproduced from Fig. 4), (magenta) orbital momentum in the film $|x| < a$ (65), (brown) spin momentum in the film (66), (black) total momentum in the film (67), (red) total orbital momentum of the SPP, (cyan) total spin momentum of the SPP.

For comparison, Fig. 7 also shows the total momentum in the film $p = p_O + p_S$ that can be presented in the form (cf. Eq. (S27))

$$p = p_{ex}^M + p_{ez}^M + p^F + p_m^M + p_{SR}^M = \frac{g}{\omega} \frac{1}{\varepsilon_2} \frac{k^2}{k_s} \left( \frac{\cosh^2 \kappa_2 x}{\cosh^2 \kappa_2 a} - \eta \frac{\kappa_1}{k^2 \varepsilon_1} \left[ \delta(x-a) + \delta(x+a) \right] \right), \; |x| < a.$$

Its integral value

$$\langle p\rangle_2 = \langle p_O\rangle_2 + \langle p_S\rangle_2 = \frac{g}{\omega}\frac{1}{\varepsilon_2 k_s}\left[\frac{k^2}{\kappa_2}\left(\frac{\kappa_2 a}{\cosh^2\kappa_2 a} - \frac{\kappa_1\varepsilon_2}{\kappa_2\varepsilon_1}\right) - 2\eta\frac{\kappa_1}{\varepsilon_1}\right] \tag{67}$$

is illustrated by the black curves in Fig. 7a, b, which, indeed, coincide with the "sums" of the brown $\langle p_O\rangle_2$ and magenta $\langle p_S\rangle_2$ curves.

For completeness, Fig. 7 contains the results for the orbital and spin momenta in dielectric taken from Fig. 4 (blue and green curves). In accord with Eqs. (41) and (66), $\langle p_S\rangle_2 = -\langle p_{DS}\rangle$, and the total spin momentum of the SPP vanishes: $\langle p_S\rangle = \langle p_S^M\rangle + \langle p_S^F\rangle + \langle p_{DS}\rangle = 0$ (cyan lines in Fig. 7), in full agreement with the general spin-momentum theory [27,15,20]. Due to the vanishing $\langle p_S\rangle$, the total momentum of the SPP $\langle p\rangle = \langle p\rangle_2 + \langle p_D\rangle$ (red curves in Fig. 7a, b) coincides with the total orbital contribution $\langle p_O\rangle$. With growing $k_s$, the red and black curves of Fig. 7 tend to the same asymptote because their difference $\langle p_D\rangle$ is relatively small and tends to zero, as is shown by the red curves in Fig. 4.

## 5. Discussion and conclusion

In this paper, a detailed analysis of the energy and momentum of the SPP waves excited in a symmetric 3-layer "insulator-metal-insulator" structure is performed. The electric and magnetic field vectors are calculated in the lossless approximation via both the phenomenological and the microscopic approach. The latter involves the motion of electrons in the metal and is based on the hydrodynamic model accounting for the quantum statistical effects. Based on the calculated field expressions, the explicit representation for the energy and momentum constituents in the dielectric and in the intermediate metal film are obtained. However, the main results concern the "integral" energy and momentum contributions obtained by the integration over the whole structure thickness.

The microscopic approach enabled us to distinguish the "field" and "material" contributions of the SPP energy and momentum in the metal film; additionally, the material contributions associated with various kinds of the electrons' motion can be identified (in the dielectric layers, the whole energy and momentum are treated as of the "field" origin). Based on the numerical example in which the dielectric layers are made of silica, and the metal can be described by the Drude model for the electron gas in silver, we analyze the behavior of the energy and momentum constituents depending on the SPP wavenumber $k_s$. The results are presented in Figs. 3 – 7 with the systematic comparison of the situations characteristic for the "high-frequency" symmetric (S) and "low-frequency" antisymmetric (AS) SPP modes.

The known physical difference between the S and AS modes is that the S mode shows the negative energy flow and the negative group velocity in the region $k_s > k_{s2}$ where $k_{s2}$ is the point of the dispersion curve maximum [28,29] (see Fig. 1). However, the most noticeable differences between the energy and momentum constituents' behavior in the S and AS modes occur at $k_s < k_{s2}$, where both modes are characterized by the "usual" positive energy flow. For example, all the energy components normalized by $|E_x(a)|^2$, where $|E_x(a)|$ is the transverse electric field amplitude at the film boundary, grow with decreasing $k_s$, and only the "field" energy in the metal film $\langle w^F\rangle_2$ of the S mode tends to zero (Fig. 3a).

Another remarkable feature can be traced from comparison of the blue curves in Figs. 6a and 6b for the material momentum component $\langle p_m^M\rangle$: in the S mode it is noticeably higher in the low-frequency region. Other differences between the curves in Figs. 4a – 7a and 4b – 7b are less perceptible, largely quantitative and, again, occur in the positive-flow region $k_s < k_{s2}$. All these differences can be associated with the difference between the dispersion curves for the S and AS modes (blue and green curves, Fig. 2), which is most apparent in the low-frequency region $k_s < k_{s2}$. In contrast, at the negative-flow region $k_s > k_{s2}$ all the momentum contributions behave quite "traditionally" both in the S and the AS modes. For example, all the momentum contributions presented in Figs. 4 – 7 are directed along the wave propagation, independently of the energy flow direction (except the spin contribution in dielectric $\langle p_{DS}\rangle$, Figs. 4, 7, and the "field" momentum in the metal $\left\langle p^F \right\rangle_2$, Fig. 5, and its spin part $\left\langle p_S^F \right\rangle_2$, Fig. 6, whose opposite direction is natural for the evanescent waves and does not affect the resulting momentum direction [7,8]). Notably, the total spin momentum in the metallic layer $\langle p_S\rangle_2$ is positive for both S and AS modes (Fig. 7).

In sum, the main differences of the momentum behavior in the S and AS modes is revealed by the "interplay" of the red and green curves in Fig. 6 which illustrate the material contributions $\left\langle p_{ex}^M \right\rangle$ (46) and $\left\langle p_{ez}^M \right\rangle$ (47) discussed in section 4.3. Coarsely speaking, their "roles" in the S and AS modes are "interchanged" but this practically does not affect the resulting material-momentum composition.

All curves presented in Figs. 4 – 7 show asymptotically linear behavior, which can be derived from the corresponding formulas taking into account that with growing $k_s$

$$\omega \to \omega_c\,, \;\; k \to \omega_c / c\,, \;\; \kappa_1 \simeq \kappa_2 \to k_s\,, \;\; \eta \to 1+\varepsilon_1\,, \;\; \varepsilon_2 \to -\varepsilon_1\,.$$

This means, for example, that the asymptotic form of the black and red curves in Fig. 7 can be described as

$$\frac{c\langle p\rangle}{\langle w\rangle} \to \frac{k_s}{k} = c\frac{k_s}{\omega_c} = \frac{k_s}{1.44\cdot 10^5\,\mathrm{cm}^{-1}}\,.$$

In application to the total SPP momentum, this result can be simply interpreted in the "quantum" spirit, assuming $\langle p\rangle \propto \hbar k_s$, $\langle w\rangle \propto \hbar\omega_c$ where $\hbar$ is the Planck constant. Figs. 4 – 7 show that situations where $c\langle p\rangle/\langle w\rangle > 1$ are rather typical both for the whole momentum and for its separate components; these illustrate the "supermomentum" concept well known for the evanescent waves and SPPs in other systems [15,16].

The results of the present paper can be considered as extension of the recently performed analysis of the dynamical characteristics in the "usual" single-interface SPP [23]; in general, they and agree with the preliminary expectations. The microscopic picture of the electrons' motion, the properties of the momentum components and details of their behaviour can be valuable in the research and applications of the SPP-induced thin-film effects, such as the induced magnetization [15–17]. The specific details of the momentum classification and distribution discussed in this paper will be useful in studies of the charge and spin dynamics in presence of the external or SPP-induced static fields, e.g., for investigation and control of the spin currents in thin-film plasmonic systems [34]. Peculiar physical actions associated with the separate momentum contributions may induce additional influences on the photo-sensitive centers in presence of high-gradient local optical fields [35], etc.

# Energy and momentum of the surface plasmon-polariton supported by a thin metal film

## Supplementary Material

A. Y. Bekshaev*[a], O. V. Angelsky[b]
[a]Physics Research Institute, I.I. Mechnikov National University, Odessa, Ukraine;
[b]Yuriy Fedkovych Chernivtsi National University, Chernivtsi, Ukraine
**bekshaev@onu.edu.ua*

Here we offer the systematic inventory of expressions describing the spatial distributions of the dynamical characteristics (energy, energy flow, momentum and spin) for the SPP field in the metallic layer (medium 2, see Fig. 1), accompanied by some numerical illustrations performed for the conditions of point B in Fig. 2. Only results for the S mode are explicitly presented; corresponding expressions for the AS mode can be obtained by the replacements (8) and (19):

$$\cosh \kappa_2 x \rightleftarrows \sinh \kappa_2 x, \ \cosh \gamma x \rightleftarrows \sinh \gamma x \tag{S1}$$

(note, however, that $\cosh 2\kappa_2 x = \sinh^2 \kappa_2 x + \cosh^2 \kappa_2 x$ and $\sinh 2\kappa_2 x = 2\sinh \kappa_2 x \cosh \kappa_2 x$ remain unchanged). The formulas below are obtained using the microscopic hydrodynamic approach based on Eqs. (1) and (10) – (18) but, in contrast to the main text, with preserving the "significant" NS terms that do not vanish in the limit (22). This means that the "insignificant" $\gamma^{-1}$-terms proportional to negative degrees of $\gamma$ are normally not kept (provided that these are not used in intermediate transformations producing "significant" NS terms, see Eq. (S23) for example). The special attention is paid to the "singular" NS terms proportional to $\gamma \cosh(\gamma x)$ or $\gamma \sinh(\gamma x)$, which describe the strictly-localized near-surface pikes whose magnitude grows infinitely upon the condition (22), and whose meaningful contributions to the integral values (23) can be expressed via the delta-function due to relations (24). However, one should remember that Eqs. (24) are just approximations, and really the "delta-function" in (24) has the finite peak value: for conditions (20) and (21),

$$\delta(0) = \gamma \tanh \gamma a \simeq \gamma = 5.78 \cdot 10^7 \ \text{cm}^{-1}. \tag{S2}$$

In particular, using the representation (24), the electron density (18) for the S mode can be presented in the form

$$en \simeq -\frac{A}{4\pi} e^{-\kappa_1 a} \frac{\eta}{\varepsilon_2} \left[ \delta(x-a) - \delta(x+a) \right] e^{ik_s z} . \tag{S3}$$

Then, the "integral" charge accumulated near the metal-dielectric interface is described by the equation

$$\int\limits_{a-0}^{a} en(x)\,dx \simeq -\frac{A}{4\pi} e^{-\kappa_1 a} \frac{\eta}{\varepsilon_2} e^{ik_s z} ; \tag{S4}$$

this is the complex amplitude of the instantaneous charge oscillations with frequency $\omega$.

### S1. Electric and magnetic fields, electron velocity and density

These distributions are given explicitly in Eqs. (15) – (18), here they are illustrated by Fig. S1. The curves represent the data normalized with respect to $A' = A\exp(-\kappa_1 a) = E_x(a)$ (see Eq. (26)); we use the Gaussian units everywhere, so the unit of $A$, $A'$ and $E_x$ is statV/cm = (g/cm)$^{1/2}$s$^{-1}$ = $3\cdot 10^4$ V/m. The curves show that the NS terms inspire rapid changes of the normal electric-field and velocity components, $E_x$ and $v_x$, in the NS region, required by the boundary conditions. The NS changes of $E_z$ and $v_z$, caused by the insignificant NS terms of the second Eqs. (16), (17), are minor (e.g., the minimum of the green curve for $E_z$ slightly to the left of the interface $x = 15$ nm). The most important feature of the field distributions is that $E_x$ changes the sign at a certain point $x = x_e$ inside the metal film (upon conditions accepted for calculations of Fig. S1, $x_e = 14.81$ nm); the phenomenologically expected behavior of $E_x$ (without the NS terms in (16)) is shown by the dashed blue line.

The curves of Fig. S1 qualitatively agree with the results obtained for the single-interface SPP [21]. Fig. S1 illustrates the situation in the upper half of the film ($0 < x < 15$ nm); for $x < 0$, according to the nature of the S mode, $H_y$, $E_x$ and $v_x$ behave symmetrically while $E_z$, $v_z$ and $n$ change the signs. This agrees with the zero values of $E_z$, $v_z$ and $n$ at $x = 0$.

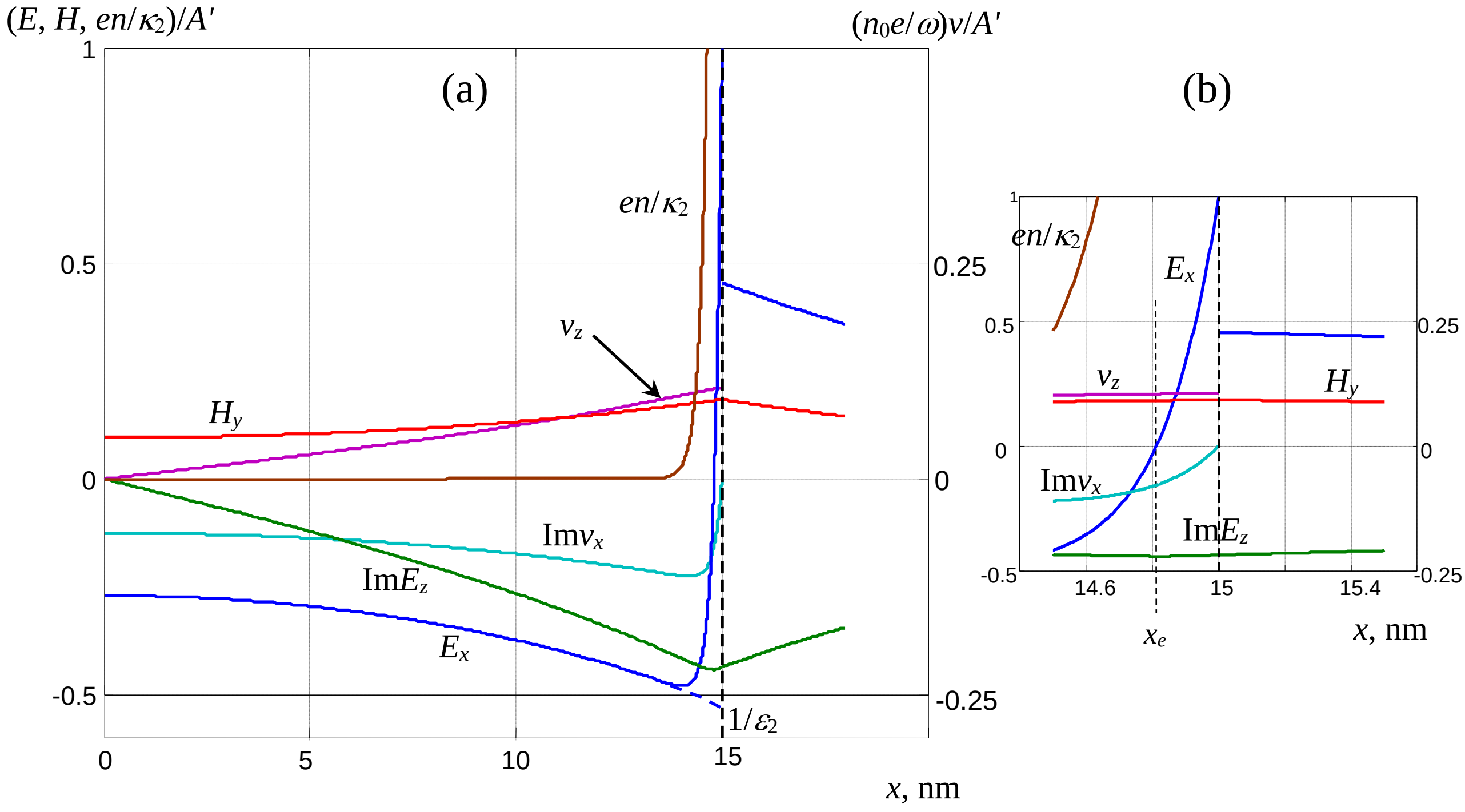

Fig. S1. Amplitudes of (blue, green) electric (3), (16) and (red) magnetic (3), (15) field components, (brown) normalized electron density (18) and (cyan, magenta, right vertical scale) normalized electron velocity components (17) calculated for the conditions (20), (21) (the SPP frequency $\omega = 4.52 \cdot 10^{15}$ s$^{-1}$). The interface plane is marked by the vertical dashed line, the panel (b) shows the NS region of (a) in a magnified horizontal scale, $x_e$ denotes the position where $E_x = 0$.

## S2. Electric polarization

The results (15) – (18) supply a complete characterization of the electron plasma in the film 2 and, particularly, describe the SPP-induced polarization. The electric polarization can be found following to the general scheme of [21]. In the monochromatic field, each electron performs an oscillatory motion whose velocity $\mathbf{v}$ is associated with the displacement $(i/\omega)\mathbf{v}$ from its “neutral” position. This displacement produces an effective dipole with a dipole moment per unit volume

$$\mathbf{d} = n_0 (ie/\omega)\mathbf{v}, \tag{S5}$$

and Eqs. (16), (17) determine the linear dependence

$$\mathbf{d} = \hat{\alpha}\mathbf{E}. \tag{S6}$$

Here $\hat{\alpha}$ is the diagonal polarizability tensor with spatially variable elements $\alpha_{xx}$, $\alpha_{zz}$. In the approximation (22), (24), the variability of $\alpha_{zz}$ can be neglected, and

$$\alpha_{zz} = -\frac{\eta}{4\pi} = \frac{\varepsilon_2 - 1}{4\pi}, \quad \alpha_{xx} = \alpha_{zz} \frac{\dfrac{\cosh \kappa_2 x}{\cosh \kappa_2 a} - \dfrac{\cosh \gamma x}{\cosh \gamma a}}{\dfrac{\cosh \kappa_2 x}{\cosh \kappa_2 a} - \eta \dfrac{\cosh \gamma x}{\cosh \gamma a}} \tag{S7}$$

(note that $\alpha_{xx}(\pm a) = 0$). Their frequency derivatives are

$$\alpha'_{xx} = \frac{d\alpha_{xx}}{d\omega} = \alpha'_{zz} \left[ 1 + \frac{(\eta - 1)\dfrac{\cosh \gamma x}{\cosh \gamma a}}{\dfrac{\cosh \kappa_2 x}{\cosh \kappa_2 a} - \eta \dfrac{\cosh \gamma x}{\cosh \gamma a}} + \eta \frac{\cosh \gamma x}{\cosh \gamma a} \frac{\dfrac{\cosh \kappa_2 x}{\cosh \kappa_2 a} - \dfrac{\cosh \gamma x}{\cosh \gamma a}}{\left( \dfrac{\cosh \kappa_2 x}{\cosh \kappa_2 a} - \eta \dfrac{\cosh \gamma x}{\cosh \gamma a} \right)^2} \right]$$

$$= \alpha'_{zz} \frac{\cosh \kappa_2 x}{\cosh \kappa_2 a} \frac{\dfrac{\cosh \kappa_2 x}{\cosh \kappa_2 a} - \dfrac{\cosh \gamma x}{\cosh \gamma a}}{\left( \dfrac{\cosh \kappa_2 x}{\cosh \kappa_2 a} - \eta \dfrac{\cosh \gamma x}{\cosh \gamma a} \right)^2}, \quad \alpha'_{xx}(\pm a) = 0; \tag{S8}$$

$$\alpha'_{zz} = \frac{d\alpha_{zz}}{d\omega} = \frac{\eta}{2\pi\omega} = \frac{1 - \varepsilon_2}{2\pi\omega} = \frac{1}{4\pi} \frac{d\varepsilon_2}{d\omega}. \tag{S9}$$

These results are used for the calculations of some momentum constituents, see Eq. (S20) below.

### S3. Energy and energy flow density

In the metallic layer 2, the energy density is determined by Eq. (27) including the “field” and material parts. The analogs of Eqs. (28), (29) with preserved significant NS terms take the forms

$$w^F = \frac{g}{2\varepsilon_2^2} \left[ \left( 1 + \varepsilon_2^2 \frac{k^2}{k_s^2} \right) \frac{\cosh^2 \kappa_2 x}{\cosh^2 \kappa_2 a} + \frac{\kappa_2^2}{k_s^2} \frac{\sinh^2 \kappa_2 x}{\cosh^2 \kappa_2 a} - 2\eta \frac{\cosh \kappa_2 x \cosh \gamma x}{\cosh \kappa_2 a \cosh \gamma a} + \eta^2 \frac{\cosh^2 \gamma x}{\cosh^2 \gamma a} \right], \tag{S10}$$

$$w^M = \frac{g\eta}{2\varepsilon_2^2} \left[ \frac{\cosh^2 \kappa_2 x}{\cosh^2 \kappa_2 a} + \frac{\kappa_2^2}{k_s^2} \frac{\sinh^2 \kappa_2 x}{\cosh^2 \kappa_2 a} - 2 \frac{\cosh \kappa_2 x \cosh \gamma x}{\cosh \kappa_2 a \cosh \gamma a} + \frac{\cosh^2 \gamma x}{\cosh^2 \gamma a} - \varepsilon_2 \frac{\sinh^2 \gamma x}{\cosh^2 \gamma a} \right]. \tag{S11}$$

The characteristic features of spatial distributions (S10), (S11) are the kinks in the close vicinity of the interface (blue, green and red curves in Fig. S2). The kinks are similar to those obtained for the “simple” SPP supported by a single metal-dielectric interface [21], and are associated with the vanishing $E_x$ at $x = x_e$.

The energy flow density in the film directly follows from the definition [21,25]

$$S = \frac{c}{8\pi} E_x^* H_y + \frac{1}{2} m\beta^2 n^* v_z \tag{S12}$$

where the second (material) term is negligible in the approximation (22), (24) [21] while Eqs. (15) and (16) give the refined Eq. (36):

$$S = g \frac{c}{\varepsilon_2} \frac{k}{k_s} \left( \frac{\cosh^2 \kappa_2 x}{\cosh^2 \kappa_2 a} - \eta \frac{\cosh \kappa_2 x \cosh \gamma x}{\cosh \kappa_2 a \cosh \gamma a} \right), \tag{S13}$$

see the magenta curve in Fig. S2. The energy flow density (S13) changes the sign at the same point $x_e$ where $E_x$ does (cf. Fig. S1b).

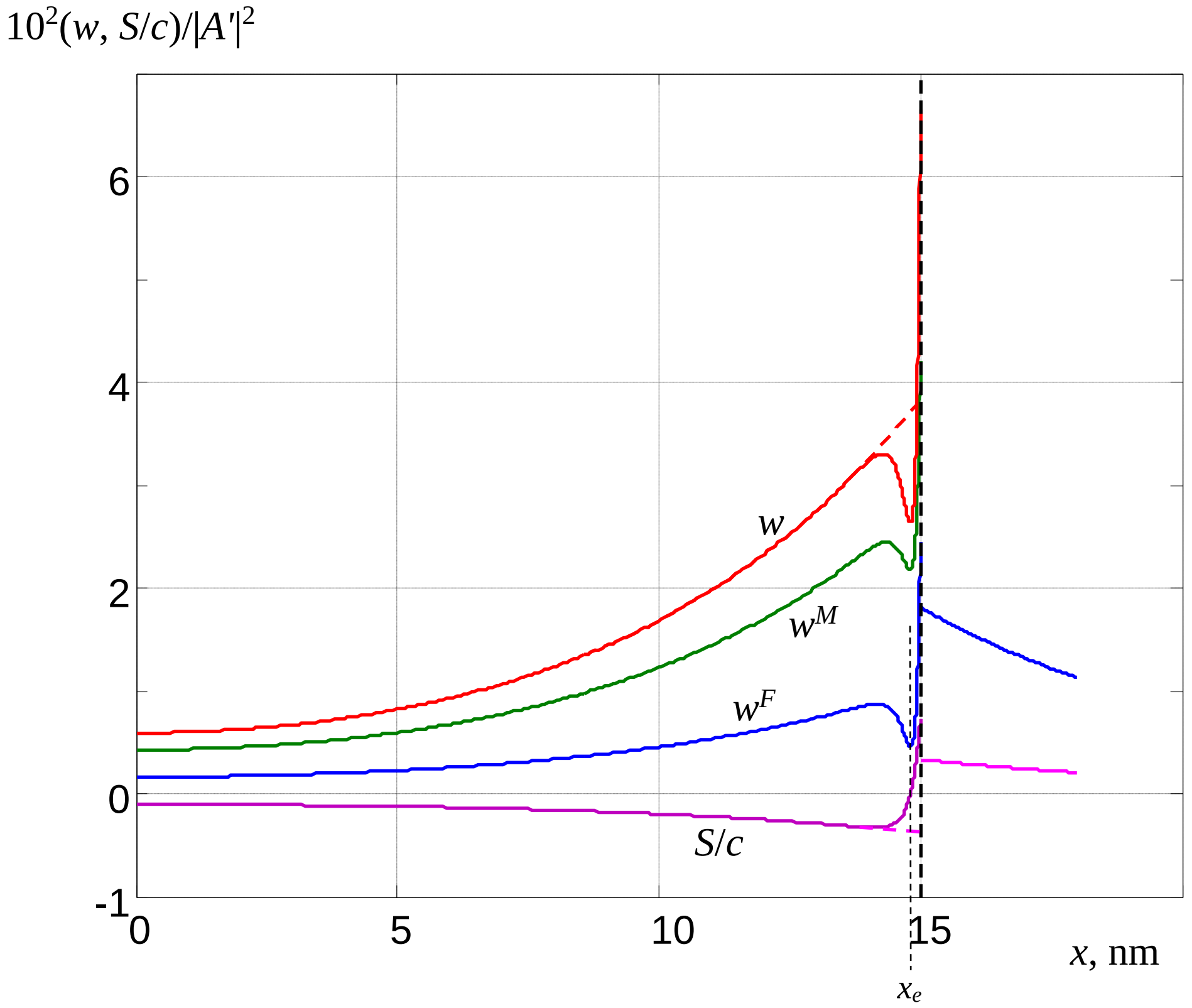

Fig. S2. Energy and energy flow density distributions calculated for the conditions (20), (21): (blue) "field" contribution (S10) and, for $x > 15$ nm, (25); (green) material contribution (S11); (red) "complete" energy density $w^F + w^M$; (magenta) energy flow density (S13) and (35). For comparison, the red and magenta dashed lines show the phenomenological results (without the NS terms). The interface plane is marked by the vertical dashed line.

## S4. "Field" and material momentum density

The "pure-field" momentum contribution can be obtained from the Poynting vector expression (34) due to relation

$$p^F = \frac{1}{c^2} S = \frac{1}{8\pi c} E_x^* H_y , \tag{S14}$$

which gives the refined expression (42):

$$p^F = \frac{g}{c} \frac{k}{\varepsilon_2 k_s} \left( \frac{\cosh^2 \kappa_2 x}{\cosh^2 \kappa_2 a} - \eta \frac{\cosh \kappa_2 x \cosh \gamma x}{\cosh \kappa_2 a \cosh \gamma a} \right), \quad p^F(\pm a) = \frac{g}{c} \frac{k}{k_s} . \tag{S15}$$

The meaning of analytical expressions describing the momentum and its separate "blocks" is illustrated by Figs. S3 – S6. Only the behavior in the upper half of the film ($x > 0$) is explicitly shown; in the region $x < 0$, all the momentum constituents are distributed symmetrically. The "field" momentum (S15) is illustrated by the blue curve in Fig. S3.

Other "blocks" of the electromagnetic momentum in the film 2 (Fig. 1) include the material contributions associated with the motion of electrons. To find these contributions, we must consider the force exerted by the field on the medium particles, which is performed in the Supplementary Document of [21], section S2, and dictates that the material contribution consists of two parts. The first one can be presented as

$$\mathbf{p}_m^M = -\frac{\eta}{8\pi c} \mathrm{Re}\left(\mathbf{E}^* \times \mathbf{H}\right) + \frac{1}{2c}\left(\frac{e\beta^2}{\omega^2} \nabla n \times \mathbf{H}\right) = \mathbf{z} p_m^M , \tag{S16}$$

where, in view of Eqs. (15), (16) and (18),

$$p_m^M = -\eta p^F + \frac{g}{c}\eta\frac{k}{k_s}\frac{\cosh\kappa_2 x\cosh\gamma x}{\cosh\kappa_2 a\cosh\gamma a} \tag{S17}$$

where $p^F$ is determined by (S15) so that finally, in view of (S7),

$$p_m^M = \frac{\alpha_{xx}}{2c}E_x^* H_y = -\frac{g}{c}\frac{\eta}{\varepsilon_2}\frac{k}{k_s}\left(\frac{\cosh^2\kappa_2 x}{\cosh^2\kappa_2 a} - \frac{\cosh\kappa_2 x\cosh\gamma x}{\cosh\kappa_2 a\cosh\gamma a}\right), \quad p_m^M(\pm a) = 0 \tag{S18}$$

(yellow curve of Fig. S3). This is an improved version of Eq. (44).

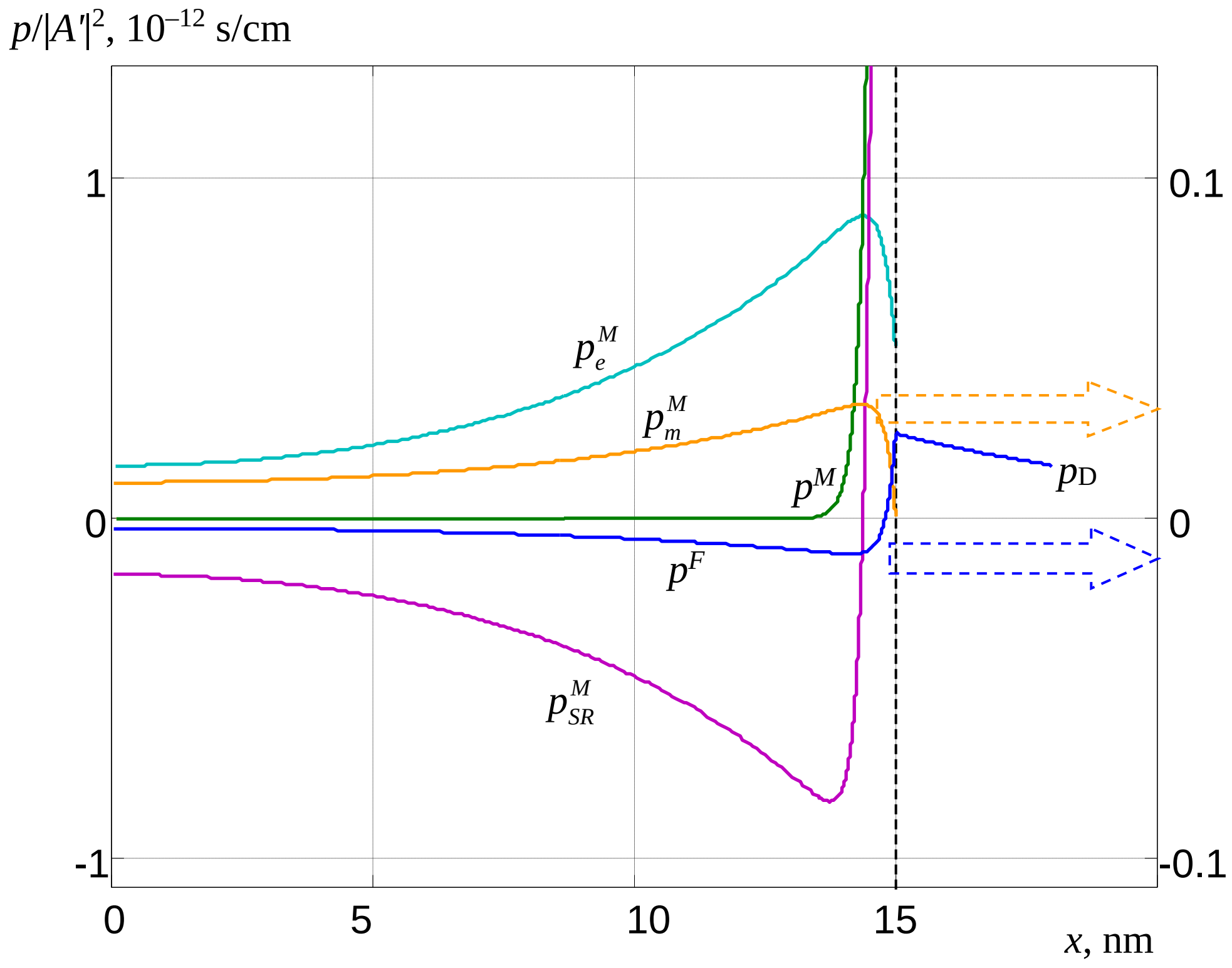

Fig. S3. SPP momentum constituents calculated for the conditions (20), (21) (point B of Fig. 2): “field” and material contributions. (Blue) “field” contribution (S15) (at $x > 15$ nm coincides with the complete momentum (39)), right vertical scale; (yellow) material Poynting momentum (S18), right vertical scale; (cyan) material “electric” contribution (S20); (magenta) “material rotational spin momentum” (S25); (green) complete material momentum (S26). The interface plane is marked by the vertical dashed line.

The whole Poynting momentum $p^P$ (which now appears in the Minkowski form) includes the field part $p^F$ (S15) and the material part $p_m^M$ (S18):

$$p^P = p^F + p_m^M = \frac{g}{c}\frac{k}{k_s}\frac{\cosh^2\kappa_2 x}{\cosh^2\kappa_2 a}. \tag{S19}$$

This quantity differs from the “genuine” electromagnetic momentum by the absence of dispersion corrections [15–17] and has no direct physical meaning. Remarkably, in (S19) the significant NS terms of $p_m^M$ and $p^F$ mutually cancel, so that the Minkowski momentum appears exactly in the same form as in the phenomenological approach (however, this correspondence can be distorted by the insignificant NS terms proportional to $\gamma^{-1}$, which are omitted in (S19)).

The second material "block" of the SPP momentum, which can be called "electric" due to its direct connection to the electric field components, follows from Eq. (S30) of [21] in the form

$$\mathbf{p}_e^M = \mathbf{z} p_e^M ,$$

$$p_e^M = p_{ex}^M + p_{ez}^M = \frac{1}{4}\mathrm{Im}\left(\alpha'_{xx} E_x^* \frac{\partial E_x}{\partial z} + \alpha'_{zz} E_z^* \frac{\partial E_z}{\partial z}\right) = \frac{k_s}{4}\left(\alpha'_{xx}\left|E_x\right|^2 + \alpha'_{zz}\left|E_z\right|^2\right). \tag{S20}$$

It can be shown that this contribution is directly associated with the dispersion corrections omitted in (S18) (see Eq. (S29) below). For the SPP field of Eqs. (16), and employing Eqs. (S8), (S9), separate summands of Eq. (S20) can be easily found:

$$p_{ex}^M = \frac{k_s}{4}\alpha'_{xx}\left|E_x\right|^2 = \frac{g}{c}\frac{\eta}{\varepsilon_2^2}\frac{k_s}{k}\left(\frac{\cosh^2 \kappa_2 x}{\cosh^2 \kappa_2 a} - \frac{\cosh \kappa_2 x \cosh \gamma x}{\cosh \kappa_2 a \cosh \gamma a}\right), \quad p_{ex}^M\left(\pm a\right) = 0;$$

$$p_{ez}^M = \frac{k_s}{4}\alpha'_{zz}\left|E_z\right|^2 = \frac{g}{\omega}\frac{\eta}{\varepsilon_2^2}\frac{\kappa_2^2}{k_s}\frac{\sinh^2 \kappa_2 x}{\cosh^2 \kappa_2 a}, \quad p_{ez}^M\left(\pm a\right) = \frac{g}{\omega}\frac{\eta}{\varepsilon_1^2}\frac{\kappa_1^2}{k_s}.$$

These equations provide refinements for the phenomenological Eqs. (46) and (47). As a result, the whole momentum "block" (S20) appears in the form

$$p_e^M = p_{ex}^M + p_{ez}^M = \frac{g}{c}\frac{\eta}{\varepsilon_2^2}\frac{k_s}{k}\left(\frac{\cosh^2 \kappa_2 x}{\cosh^2 \kappa_2 a} + \frac{\kappa_2^2}{k_s^2}\frac{\sinh^2 \kappa_2 x}{\cosh^2 \kappa_2 a} - \frac{\cosh \kappa_2 x \cosh \gamma x}{\cosh \kappa_2 a \cosh \gamma a}\right), \tag{S21}$$

$$p_e^M\left(\pm a\right) = \frac{g}{c}\frac{\eta}{\varepsilon_1^2}\frac{\kappa_1^2}{k k_s}$$

(cyan curve in Fig. S3).

Yet another material contribution originates from the rotational motion of electrons. Its calculation starts from the "material rotational spin" [15–17,21]:

$$\mathbf{s}_R^M = \frac{m n_0}{2\omega}\mathrm{Im}\left(\mathbf{v}^* \times \mathbf{v}\right) = \mathbf{y} s_R^M . \tag{S22}$$

which, via Eq. (17), entails

$$s_R^M = -\frac{2g}{\omega}\frac{\eta}{\varepsilon_2^2}\frac{\kappa_2}{k_s}\left(\frac{\cosh \kappa_2 x \sinh \kappa_2 x}{\cosh^2 \kappa_2 a}\right.$$

$$\left. - \frac{\sinh \kappa_2 x \cosh \gamma x}{\cosh \kappa_2 a \cosh \gamma a} - \frac{k_s^2}{\gamma \kappa_2}\frac{\cosh \kappa_2 x \sinh \gamma x}{\cosh \kappa_2 a \cosh \gamma a} + \frac{k_s^2}{\gamma \kappa_2}\frac{\sinh \gamma x \cosh \gamma x}{\cosh^2 \gamma a}\right). \tag{S23}$$

According to the general rule [20,21]

$$\mathbf{p}_S = \frac{1}{2}\nabla \times \mathbf{s}, \tag{S24}$$

every summand of the spin density is the source of the corresponding spin-momentum density. In this view, Eqs. (S22) and (S23) (where the $\gamma^{-1}$-terms are kept because they give meaningful contributions after subsequent transformations) determine the "material rotational spin momentum" in the form $\mathbf{p}_{SR}^M = \mathbf{z} p_{SR}^M$ with

$$p_{SR}^M = \frac{1}{2}\frac{d s_R^M}{dx} = -\frac{g}{\omega}\frac{\eta}{\varepsilon_2^2}\frac{\kappa_2}{k_s}\left\{\kappa_2 \frac{\cosh 2\kappa_2 x}{\cosh^2 \kappa_2 a} - \kappa_2\left(1 + \frac{k_s^2}{\kappa_2^2}\right)\frac{\cosh \kappa_2 x \cosh \gamma x}{\cosh \kappa_2 a \cosh \gamma a}\right.$$

$$\left. + \frac{k_s^2}{\kappa_2}\frac{\cosh 2\gamma x}{\cosh^2 \gamma a} - \tanh \kappa_2 a\left[\delta\left(x - a\right) + \delta\left(x + a\right)\right]\right\}$$

$$= -\frac{g}{\omega}\frac{\eta}{\varepsilon_2^2}k_s\left\{\frac{\kappa_2^2}{k_s^2}\frac{\cosh 2\kappa_2 x}{\cosh^2 \kappa_2 a} - \left(\frac{\kappa_2^2}{k_s^2} + 1\right)\frac{\cosh \kappa_2 x \cosh \gamma x}{\cosh \kappa_2 a \cosh \gamma a}\right.$$

$$+\frac{\cosh 2\gamma x}{\cosh^2 \gamma a}+\frac{\kappa_1\varepsilon_2}{k_s^2\varepsilon_1}\left[\delta(x-a)+\delta(x+a)\right]\Bigg\} \tag{S25}$$

where the asymptotic relation (24) has been used (see the magenta curve in Fig. S3). The first and last summands of Eq. (S25) correspond to Eqs. (51) and (52) where the "middle" terms of (S25) are omitted. The expected boundary value of the "volume" part of $p_{SR}^M$ (given by the first summand in brackets of (S25)) is

$$\left(p_{SR}^M\right)_{\text{Vol}}(a)=-\frac{g}{\omega}\frac{\eta}{\varepsilon_2^2}\frac{\kappa_2^2}{k_s}\left(1+\tanh^2\kappa_2 a\right).$$

The complete material contribution $p^M = p_e^M + p_m^M + p_{SR}^M$ (cf. Eq. (54)) reduces to the "singular" and NS terms (i.e., "volume" material momentum vanishes):

$$p^M = \frac{g}{\omega}\frac{\eta}{\varepsilon_2^2}k_s\left\{\frac{\cosh\kappa_2 x\cosh\gamma x}{\cosh\kappa_2 a\cosh\gamma a}-\frac{\cosh 2\gamma x}{\cosh^2\gamma a}-\frac{\kappa_1\varepsilon_2}{k_s^2\varepsilon_1}\left[\delta(x-a)+\delta(x+a)\right]\right\}. \tag{S26}$$

That is why the corresponding green curve in Fig. S3 practically coincides with the horizontal axis everywhere but shows a near-vertical ramp in the interface vicinity. At last, the "overall" electromagnetic momentum $p$ in the medium 2 appears as a sum of the pure-field contribution (S15) and the "singular" momentum (S26):

$$p = p^F + p^M = \frac{g}{c}\frac{1}{\varepsilon_2}\frac{k}{k_s}$$

$$\times\left\{\frac{\cosh^2\kappa_2 x}{\cosh^2\kappa_2 a}+\frac{\eta\kappa_2^2}{\varepsilon_2 k^2}\frac{\cosh\kappa_2 x\cosh\gamma x}{\cosh\kappa_2 a\cosh\gamma a}-\frac{\eta k_s^2}{\varepsilon_2 k^2}\frac{\cosh 2\gamma x}{\cosh^2\gamma a}-\frac{\eta\kappa_1}{\varepsilon_1 k^2}\left[\delta(x-a)+\delta(x+a)\right]\right\}. \tag{S27}$$

Noteworthy, both the "volume" and "singular" terms perfectly agree with the phenomenological expression (cf. Eq. (S23) of [21]), and only the NS terms allow for refinements inspired by the boundary conditions for $E_x$ and $v_x$ in the microscopic theory. The "singular" delta-term is exactly due to the jump of the phenomenological spin at the interface (see Section S6 below, and the NS behavior of the blue and magenta curves in Fig. S6).

According to (S2), despite the sharp growth in the close vicinity of the interface, the "true" boundary value of the "singular" spin momentum is finite:

$$p(a)\simeq p_M(a)\simeq p_{SR}^M(a)\simeq-\frac{g}{c}\frac{\gamma\kappa_1}{kk_s}\frac{1-\varepsilon_2}{\varepsilon_1\varepsilon_2}. \tag{S28}$$

## S5. Spin-orbital momentum decomposition

In the above Section S4, the separation of the material and "field" contributions has been considered. Now we perform the spin-orbital momentum decomposition [3,26,27] within the microscopic framework. To this end, we separately inspect each constituent of the field momentum calculated in Section S4,

$$p = p_e^M + p_{SR}^M + p^F + p_m^M .$$

Attribution of the two first terms is quite evident. One can easily see that the expression (S20), with allowance for (S8) and (S9) and discarding the NS terms, can be represented as

$$\mathbf{z}p_e^M = \frac{1}{4}\mathbf{z}\,\text{Im}\left(\alpha'_{xx}E_x^*\frac{\partial E_x}{\partial z}+\alpha'_{zz}E_z^*\frac{\partial E_z}{\partial z}\right)\simeq\frac{1}{16\pi}\frac{d\varepsilon_2}{d\omega}\text{Im}\left[\mathbf{E}^*\cdot(\nabla)\mathbf{E}\right]; \tag{S29}$$

that is, $p_e^M$ implies the dispersion correction of the orbital momentum in medium 2 [15,16,21] and is therefore its part. On the other hand, $p_{SR}^M$, originating from the spin (S22), (S23) belongs to the spin momentum "by the definition".

Other terms need a more careful investigation. Regarding the pure-field momentum $p^F$, its decomposition immediately follows from the formal decomposition of the Poynting momentum (S14) in presence of charges and currents, Eqs. (55) – (59). Then, with preserving the significant NS terms, for the spin momentum, instead of (60), one obtains

$$p_S^F = \frac{g}{2\omega}\frac{k_s}{\varepsilon_2^2}\left[-\frac{\kappa_2^2}{k_s^2}\frac{\cosh 2\kappa_2 x}{\cosh^2\kappa_2 a} + \eta\left(2 - \varepsilon_2\frac{k^2}{k_s^2}\right)\frac{\cosh\kappa_2 x\cosh\gamma x}{\cosh\kappa_2 a\cosh\gamma a} - \eta^2\frac{\cosh^2\gamma x}{\cosh^2\gamma a}\right], \tag{S30}$$

and for the orbital (canonical) part, instead of (61),

$$p_O^F = \frac{g}{2\omega}\frac{k_s}{\varepsilon_2^2}\left[\frac{\kappa_2^2}{k_s^2}\frac{\cosh 2\kappa_2 x}{\cosh^2\kappa_2 a} + 2\varepsilon_2\frac{k^2}{k_s^2}\frac{\cosh^2\kappa_2 x}{\cosh^2\kappa_2 a} - \eta\left(2 + \varepsilon_2\frac{k^2}{k_s^2}\right)\frac{\cosh\kappa_2 x\cosh\gamma x}{\cosh\kappa_2 a\cosh\gamma a} + \eta^2\frac{\cosh^2\gamma x}{\cosh^2\gamma a}\right]. \tag{S31}$$

It can be easily verified that the sum of (S30) and (S31) coincides with (S15); the boundary values of the quantities (S30) and (S31) are

$$p_S^F(a) = \frac{g}{2\omega}k_s\left[\frac{k^2}{k_s^2}\left(1+\frac{1}{\varepsilon_1}\right) - 1 - \frac{1}{\varepsilon_1^2}\right],\quad p_O^F(a) = \frac{g}{2\omega}k_s\left[\frac{k^2}{k_s^2}\left(1-\frac{1}{\varepsilon_1}\right) + 1 + \frac{1}{\varepsilon_1^2}\right]. \tag{S32}$$

In turn, the decomposition of $p_m^M$ can be performed quite similarly due to relation (S17) connecting it to the "field" momentum $p^F$. The only problem that remains is to separate properly the spin and orbital parts of the additional NS summand in (S17), which is solved by the natural requirement for both parts to vanish at the interface [21]. As a result, we have $p_m^M = p_{mO}^M + p_{mS}^M$ where

$$p_{mS}^M = -\eta p_S^F + \eta p_S^F(a)\frac{\cosh\kappa_2 x\cosh\gamma x}{\cosh\kappa_2 a\cosh\gamma a} = -\eta p_S^F + \frac{g}{2\omega}\eta k_s\left[\frac{k^2}{k_s^2}\left(1+\frac{1}{\varepsilon_1}\right) - 1 - \frac{1}{\varepsilon_1^2}\right]\frac{\cosh\kappa_2 x\cosh\gamma x}{\cosh\kappa_2 a\cosh\gamma a}$$

$$= \frac{g}{2\omega}\eta\frac{k_s}{\varepsilon_2^2}\left\{\frac{\kappa_2^2}{k_s^2}\frac{\cosh 2\kappa_2 x}{\cosh^2\kappa_2 a} + \left[\frac{k^2}{k_s^2}\left(\varepsilon_2 + \frac{\varepsilon_2^2}{\varepsilon_1}\right) - \eta^2 - 1 - \frac{\varepsilon_2^2}{\varepsilon_1^2}\right]\frac{\cosh\kappa_2 x\cosh\gamma x}{\cosh\kappa_2 a\cosh\gamma a} + \eta^2\frac{\cosh^2\gamma x}{\cosh^2\gamma a}\right\}, \tag{S33}$$

$$p_{mO}^M = -\eta p_O^F + \eta p_O^F(a)\frac{\cosh\kappa_2 x\cosh\gamma x}{\cosh\kappa_2 a\cosh\gamma a} = -\eta p_O^F + \frac{g}{2\omega}\eta k_s\left[\frac{k^2}{k_s^2}\left(1-\frac{1}{\varepsilon_1}\right) + 1 + \frac{1}{\varepsilon_1^2}\right]\frac{\cosh\kappa_2 x\cosh\gamma x}{\cosh\kappa_2 a\cosh\gamma a}$$

$$= \frac{g}{2\omega}\eta\frac{k_s}{\varepsilon_2^2}\left\{-\frac{\kappa_2^2}{k_s^2}\frac{\cosh 2\kappa_2 x}{\cosh^2\kappa_2 a} - 2\varepsilon_2\frac{k^2}{k_s^2}\frac{\cosh^2\kappa_2 x}{\cosh^2\kappa_2 a}\right.$$

$$\left. + \left[\frac{k^2}{k_s^2}\left(\varepsilon_2 - \frac{\varepsilon_2^2}{\varepsilon_1}\right) + 1 + \eta^2 + \frac{\varepsilon_2^2}{\varepsilon_1^2}\right]\frac{\cosh\kappa_2 x\cosh\gamma x}{\cosh\kappa_2 a\cosh\gamma a} - \eta^2\frac{\cosh^2\gamma x}{\cosh^2\gamma a}\right\}. \tag{S34}$$

The spatial dependencies described by Eqs. (S30) – (S34) are presented in Fig. S4 by the red, green, magenta and cyan curves. For comparison, the complete "field" and material contributions are also shown in Fig. S4 by the blue and yellow curves taken from Fig. S3 (where they were presented in a magnified vertical scale).

Putting together (S33) and (S25), we can calculate the total material contribution to the spin momentum $p_S^M = p_{mS}^M + p_{SR}^M$:

$$p_S^M = -\frac{g}{2\omega}\eta\frac{k_s}{\varepsilon_2^2}\left\{\frac{\kappa_2^2}{k_s^2}\frac{\cosh 2\kappa_2 x}{\cosh^2\kappa_2 a} - \left[3 - \eta^2 - \frac{\varepsilon_2^2}{\varepsilon_1^2} + \frac{k^2}{k_s^2}\varepsilon_2^2\left(\frac{1}{\varepsilon_1} - \frac{1}{\varepsilon_2}\right)\right]\frac{\cosh\kappa_2 x\cosh\gamma x}{\cosh\kappa_2 a\cosh\gamma a}\right.$$

$$\left. + 2\frac{\cosh 2\gamma x}{\cosh^2\gamma a} - \eta^2\frac{\cosh^2\gamma x}{\cosh^2\gamma a} + \frac{2\kappa_1\varepsilon_2}{k_s^2\varepsilon_1}\left[\delta(x-a) + \delta(x+a)\right]\right\} \tag{S35}$$

and the material contribution to the orbital momentum $p_O^M = p_{mO}^M + p_e^M$:

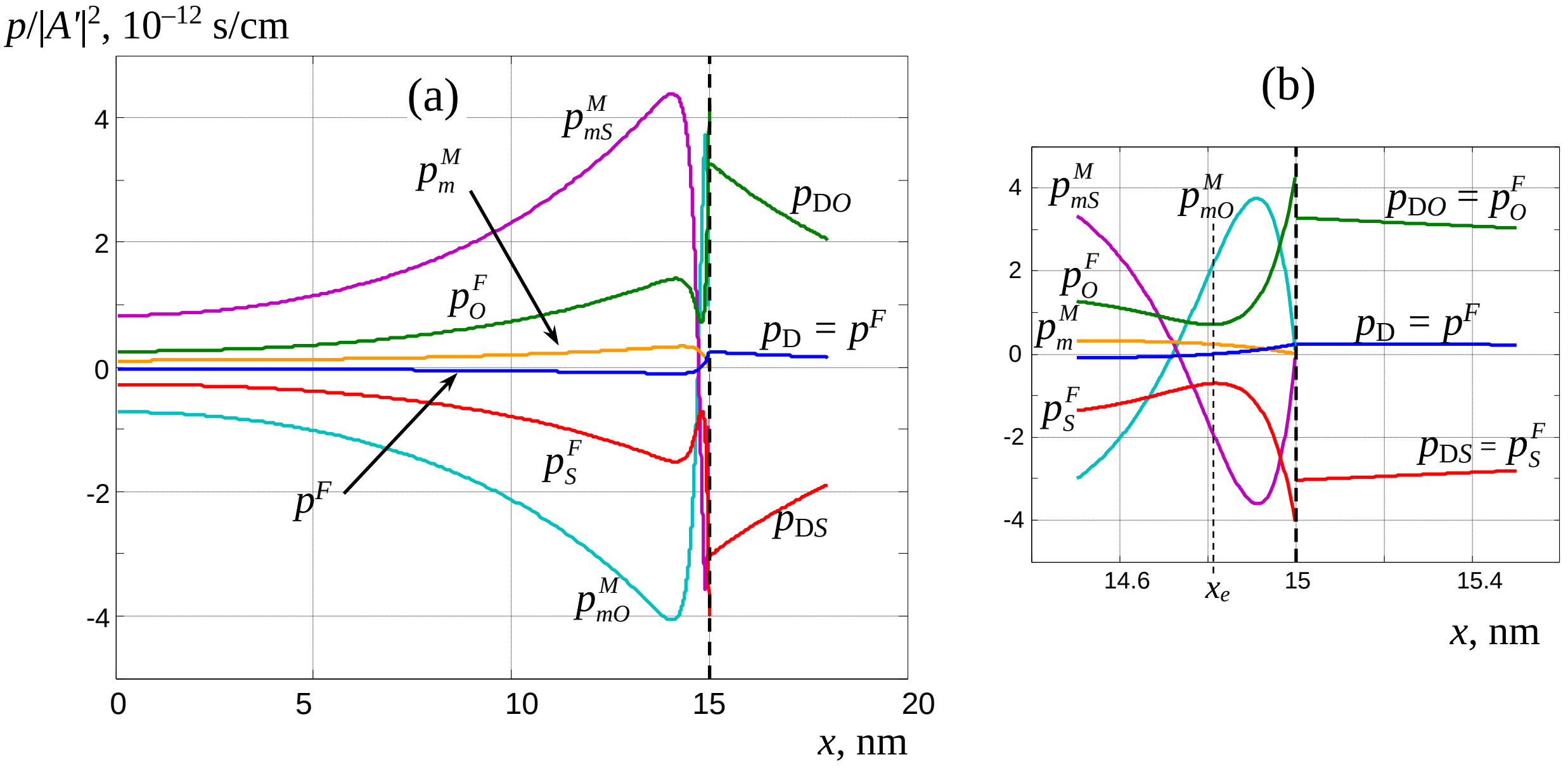

Fig. S4. SPP momentum constituents calculated for the conditions (20), (21): Spin-orbital decomposition of the “field” (S15) and material (S17) momentum. (Blue) complete “field” momentum (S15) and (39) (for $x > 15$ nm); (green) orbital “field” momentum (S31) and (40) (for $x > 15$ nm); (red) spin “field” momentum (S30) and (40) (for $x > 15$ nm); (cyan) orbital material contribution (S34); (magenta) spin material contribution (S33); (yellow) material Poynting contribution (S17). Blue and yellow curves are the same as in Fig. S3. The interface plane is marked by the vertical dashed line, the panel (b) shows the NS region of (a) in a magnified horizontal scale.

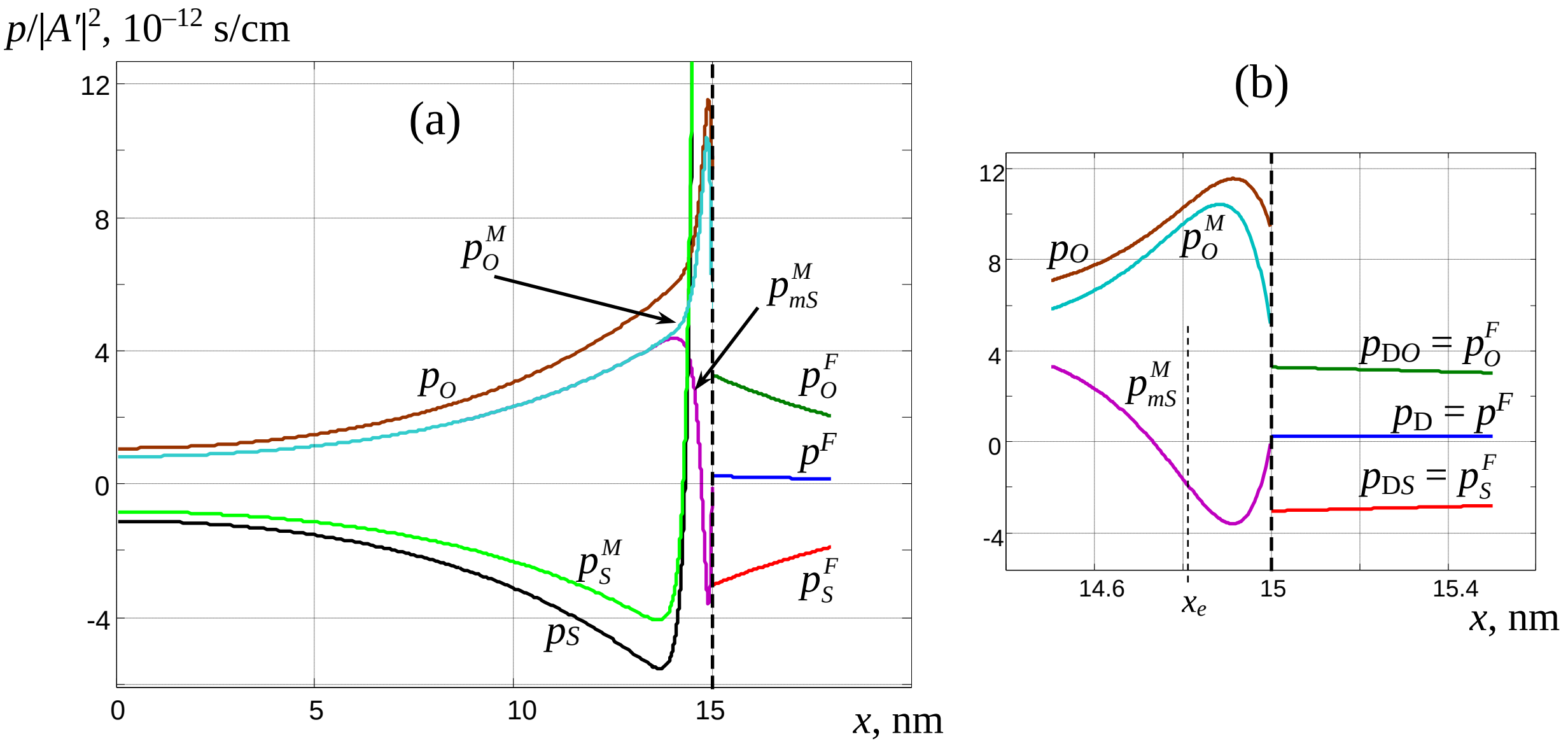

Fig. S5. SPP momentum constituents calculated for the conditions (20), (21): Complete spin and orbital contributions and their material ingredients. (Blue) complete “field” momentum (39) (for $x > 15$ nm); (green) orbital “field” momentum (40) (for $x > 15$ nm); (red) spin “field” momentum (40) (for $x > 15$ nm); (brown) complete orbital momentum $p_O = p_O^M + p_O^F$ (S31), (S36); (black) complete spin momentum $p_S = p_S^M + p_S^F$ (S30), (S35); (cyan) orbital material momenum (S36); (light-green) spin material momentum (S35); (magenta) spin material contribution (S33) (the same as in Fig. S4). The interface plane is marked by the vertical dashed line, the panel (b) shows the NS region of (a) in a magnified horizontal scale.

$$p_O^M = \frac{g}{2\omega}\eta\frac{k_s}{\varepsilon_2^2}\left\{\frac{\kappa_2^2}{k_s^2}\frac{\cosh 2\kappa_2 x}{\cosh^2\kappa_2 a} - \left[1-\eta^2-\frac{\varepsilon_2^2}{\varepsilon_1^2}+\frac{k^2}{k_s^2}\varepsilon_2^2\left(\frac{1}{\varepsilon_1}-\frac{1}{\varepsilon_2}\right)\right]\frac{\cosh\kappa_2 x\cosh\gamma x}{\cosh\kappa_2 a\cosh\gamma a} - \eta^2\frac{\cosh^2\gamma x}{\cosh^2\gamma a}\right\}. \quad \text{(S36)}$$

These quantities are illustrated by the light-green and cyan curves of Fig. S5. Note that the sum $p_O^M + p_S^M$ of Eqs. (S35) and (S36) expectedly agrees with the complete material momentum (S26). Interestingly, the "volume" parts of $p_O^M$ and $p_{mS}^M$ (first terms of expressions (S36) and (S33)) coincide, and the cyan and magenta curves merge outside the NS region, which seems to be an occasional property not dictated by any general arguments.

Finally, the complete spin momentum of the field, $p_S = p_S^M + p_S^F$, is built by combining the "field" (S30) and material (S35) contributions, as well as the orbital momentum $p_O = p_O^M + p_O^F$ "unites" the results (S31) and (S36). Their spatial distributions are illustrated by the black and brown curves in Fig. S5.

### S6. Spin density of the SPP

Spin in the dielectric layers ($|x| > a$) is determined after the Minkowski paradigm [20,21],

$$\mathbf{s}_{\mathrm{D}} = \frac{\varepsilon_1}{16\pi\omega}\mathrm{Im}\left(\mathbf{E}^*\times\mathbf{E}\right) = \frac{\varepsilon_1}{16\pi\omega}\mathbf{y}\,\mathrm{Im}\left(E_z^*E_x - E_x^*E_z\right) = \mathbf{y}s_{\mathrm{D}}, \quad \text{(S37)}$$

$$s_{\mathrm{D}} = \mathrm{sgn}\left(x\right)\frac{g}{\omega}\frac{\kappa_1}{\varepsilon_1 k_s}e^{-2\kappa_1\left(|x|-a\right)}, \quad s_{\mathrm{D}}\left(\pm a\right) = \pm\frac{g}{\omega}\frac{\kappa_1}{\varepsilon_1 k_s}, \quad \left\langle s_{\mathrm{D}}\right\rangle = 0 \quad (|x| > a) \quad \text{(S38)}$$

(blue curve at $x > 15$ nm in Fig. S6).

For the microscopic spin calculation in the metallic layer ($|x| < a$), a natural way is formally putting together different contributions associated with the rotational motion of the field vectors and field-driven material particles [23,15,16] – in the same manner that was applied considering the momentum contributions in Section S4. One of the spin contributions has already been mentioned: it is the material spin associated with the rotational motion of electrons (S22), (S23), which was employed as an auxiliary item for the momentum analysis. Here we reproduce this contribution in the final form with omitted "insignificant" $\gamma^{-1}$-terms:

$$s_R^M = -\frac{2g}{\omega}\frac{\eta}{\varepsilon_2^2}\frac{\kappa_2}{k_s}\left(\frac{\cosh\kappa_2 x\sinh\kappa_2 x}{\cosh^2\kappa_2 a} - \frac{\sinh\kappa_2 x\cosh\gamma x}{\cosh\kappa_2 a\cosh\gamma a}\right). \quad \text{(S39)}$$

Its behavior is illustrated by the cyan curve in Fig. S6. The rapid growth of this spin in the NS region is responsible for the "singular" spin momentum (S25).

Another spin contribution is the "field" spin which formally follows from Eq. (22) of [20] applied to the microscopic TM field in medium 2 with $\varepsilon = \mu = 1$ and obeying Eqs. (10). However, the "formal" spin definition used in [20] was obtained for the fields without free charges, and its non-critical use may miss some medium-associated corrections that were already applied for the momentum calculations above. To avoid this, the corresponding spin contributions will be calculated based on the relation (S24) between the spin and spin momentum due to which, for the SPP geometry, any **z**-oriented spin momentum "block" is related to the associated **y**-directed spin contribution $s$, $p_S = (1/2)(ds/dx)$. Hence, the spin expression can be recovered via integration of $p_S(x)$ with the natural boundary condition $s\left(0\right) = 0$ (in the symmetric thin-film structure of Fig. 1, all the SPP momentum distributions considered in Sections 4 and 5 are even with respect to $x = 0$ while the spin density is an odd function of $x$). Then, based on the results for $p_S^F$ (S30) and $p_{mS}^M$ (S33), we obtain the corresponding "field" and material "blocks" of the spin density in medium 2 (terms proportional to $\gamma^{-1}$ are omitted as usual):

$$s^F = -\frac{g}{2\omega}\frac{\kappa_2}{\varepsilon_2^2 k_s}\frac{\sinh 2\kappa_2 x}{\cosh^2 \kappa_2 a}, \quad s_m^M = -\eta s^F = \frac{g}{2\omega}\eta\frac{\kappa_2}{\varepsilon_2^2 k_s}\frac{\sinh 2\kappa_2 x}{\cosh^2 \kappa_2 a}, \tag{S40}$$

$$s^F(\pm a) = -\frac{1}{\eta}s_m^M(\pm a) = \mp\frac{g}{\omega}\frac{\kappa_2}{\varepsilon_2^2 k_s}\tanh\kappa_2 a = \pm\frac{g}{\omega}\frac{\kappa_1}{\varepsilon_2\varepsilon_1 k_s}$$

(see the blue and green curves in Fig. S6).

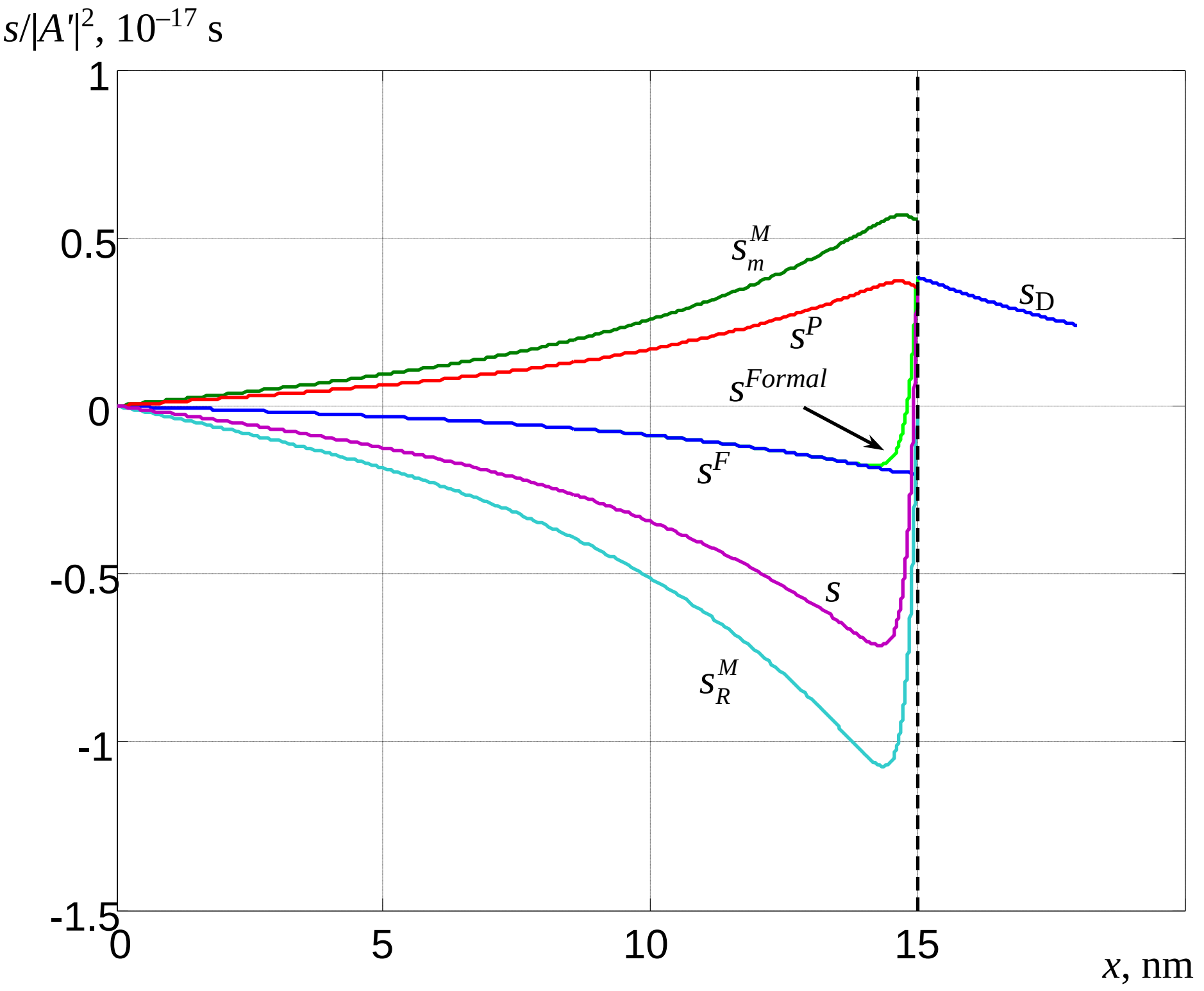

Fig. S6. SPP spin constituents calculated for the conditions (20), (21). (Blue) “field” contribution (S38) (for $x > 15$ nm) and (S40) originating from $p_S^F$ (S30); (light green) “formal” “field” spin (S41); (green) material contribution (S40) originating from $p_{mS}^M$ (S33); (red) “naïve” Minkowski spin (S42); (cyan) “rotational” material spin (S22), (S39); (magenta) complete spin (S43). The interface plane is marked by the vertical dashed line. In the region $x < 0$ (not shown) all the spin components behave antisymmetrically, $s_\alpha^\beta(-x) = -s_\alpha^\beta(x)$, so their integral values vanish.

It is instructive to compare this result with the “formal spin” derived from the formal spin definition [20]

$$s^{Formal} = \frac{1}{16\pi\omega}\mathbf{y}\cdot\mathrm{Im}\left(\mathbf{E}^*\times\mathbf{E}\right) = s^F + \frac{g}{\omega}\frac{\kappa_2}{k_s}\frac{\eta}{\varepsilon_2^2}\frac{\sinh\kappa_2 x\cosh\gamma x}{\cosh\kappa_2 a\cosh\gamma a} \tag{S41}$$

which is illustrated by the light-green curve in Fig. S6. Just like in the single-interface SPP [21], $s^{Formal}$ only contains the “field” spin $s^F$ (however, modified by the NS term), without the specific material contribution $s_m^M$. Besides, both expressions (S40) are free from significant NS terms and show the smooth “volume” variation up to the interface while the spin $s^{Formal}$ rapidly changes near the interface, reversing the sign together with $E_x$ (16). The NS term in (S41), responsible for this difference, appears because the term $\sim\mathrm{Im}\left(\mathbf{E}^* n\right)$ of (55), taken into account in (S30) and, consequently, in (S40), is omitted in $s^{Formal}$ (S41). As a result, $s^F$ is discontinuous at $x = a$ while $s^{Formal}$ is continuous with the spin in dielectric.

Interestingly, the quantity

$$s^P = s^F + s_m^M = -\frac{g}{2\omega}\frac{\kappa_2}{\varepsilon_2 k_s}\frac{\sinh 2\kappa_2 x}{\cosh^2 \kappa_2 a} \tag{S42}$$

visually coincides with the "naïve" Minkowski spin in the film described by the standard spin equations without the dispersion corrections ($\tilde{\varepsilon}_2 = \varepsilon_2$ in Eq. (S13) of the Supplementary Document of [21]). One can immediately persuade that $s^P(\pm a) = s_{\mathrm{D}}(\pm a)$ (see (S38)), which agrees with the fact that the Minkowski spin is continuous [20]. However, the "genuine" value of the sum $s^P = s^F + s_m^M$ differs from the right-hand side of (S42) by the omitted insignificant NS terms, for which reason the curves for $s^P$ and $s_{\mathrm{D}}$ slightly mismatch at the interface $x = 15$ nm (Fig. S6).

The complete spin in medium 2 is determined by equation

$$s = s^F + s_m^M + s_R^M = -\frac{g}{\omega}\frac{\kappa_2}{\varepsilon_2^2 k_s}\left[(1+\eta)\frac{\cosh\kappa_2 x \sinh\kappa_2 x}{\cosh^2 \kappa_2 a} - 2\eta\frac{\sinh\kappa_2 x \cosh\gamma x}{\cosh\kappa_2 a \cosh\gamma a}\right]. \tag{S43}$$

Due to its NS term, the overall microscopic spin (S43) is continuous, $s(\pm a) = s_{\mathrm{D}}(\pm a)$ (cf. Eq. (S38)): in Fig. S6, the magenta curve approaching the interface plane at $x < 15$ nm matches the blue line approaching the boundary from $x > 15$ nm.